\documentclass[aps,a4paper,superscriptaddress,twocolumn]{revtex4-1}
\usepackage{tensor}
\usepackage[utf8]{inputenc}
\usepackage{graphicx}
\graphicspath{ {images/} }
\usepackage{amsmath}
\usepackage{amssymb}
\usepackage{enumerate}
\usepackage{subfigure}
\usepackage{tabularx}
\usepackage{lipsum} 
\usepackage{float} 
\usepackage[colorlinks=true, pdfstartview=FitV, linkcolor=blue, citecolor=red, urlcolor=black, breaklinks=true]{hyperref}
\newcommand{\be}{\begin{equation}}
\newcommand{\ee}{\end{equation}}
\newcommand{\ben}{\begin{eqnarray}}
\newcommand{\een}{\end{eqnarray}}
\newcommand{\bes}{\begin{subequations}}
\newcommand{\ees}{\end{subequations}}
\def\bal#1\eal{\begin{align}#1\end{align}}

\newcommand{\LL}{{\mathcal L}}

\newcommand{\veps}{\varepsilon}
\newcommand{\qt}[1]{``#1''}
\newcommand{\pd}[2]{\ensuremath{\frac{\partial#1}{\partial#2}}}

\newcommand{\pb}[1]{\ensuremath{\partial_{#1}}}

\begin{document}
\title{Multimagnetic Monopoles}

\author{D. Bazeia}\affiliation{Departamento de F\'\i sica, Universidade Federal da Para\'\i ba, 58051-970 Jo\~ao Pessoa, PB, Brazil}
\author{M. A. Liao}\affiliation{Departamento de F\'\i sica, Universidade Federal da Para\'\i ba, 58051-970 Jo\~ao Pessoa, PB, Brazil}
\author{M. A. Marques}\affiliation{Departamento de Biotecnologia, Universidade Federal da Para\'iba, 58051-900 Jo\~ao Pessoa, PB, Brazil}\affiliation{Departamento de F\'\i sica, Universidade Federal da Para\'\i ba, 58051-970 Jo\~ao Pessoa, PB, Brazil}
\begin{abstract}
In this work we investigate the presence of magnetic monopoles that engender multimagnetic structures, which arise from an appropriate extension of the $\rm{SU(2)}$ gauge group. The investigation is based on a modified relativistic theory that contain several gauge and matter fields, leading to a Bogomol'nyi bound and thus to a first order framework, from which stable multimagnetic solutions can be constructed. We illustrate our findings with several examples of stable magnetic monopoles with multimagnetic properties.    
\end{abstract}

\date{\today}
\maketitle

\section{Introduction}
Magnetic monopole has been an important object of investigation in theoretical physics ever since Dirac first showed that it can be incorporated into the framework of electromagnetism in a simple and elegant way~\cite{Dirac, DiracII}. Of particular importance is the demonstration, also given by Dirac, that their mere existence could, when combined with the postulates of quantum mechanics, provide a natural explanation for the quantization of electric (and magnetic) charge.

Despite this theoretical triumph, lack of experimental evidence for the existence of monopoles led Dirac's theory to be regarded as a mathematical curiosity. Almost fifty years later,  't Hooft~\cite{thooft} and Polyakov~\cite{polyakov} independently found a topological monopole solution for a Yang-Mills-Higgs theory; see also Refs. \cite{ps,bogo,AA}. The 't Hooft-Polyakov monopole presents a magnetic charge which is directly proportional to a topological invariant called the winding number, or topological degree~\cite{BB,manton}.  Perhaps even more important than the solution itself was the realization that magnetic monopoles were an unavoidable consequence of spontaneous symmetry breaking in grand unification theories~\cite{thooft}. The so called monopole problem, regarding the apparent absence of these particles in our universe despite the theoretical predictions of their production in the early universe~\cite{Einhorn} was one of the main motivations for  the development of inflationary universe models~\cite{Guth, inflation}.

The study of magnetic monopoles has further evolved by the introduction of models with enhanced gauge symmetry~\cite{Witten, Shifman}. This enhancement allows for more complex monopoles, which now emerge from a much richer topology and present internal structure.  Works~\cite{internal, smallandhollow, bimag} have been developed along these lines, and we have also conducted a similar investigation in (2,1) dimensions, which led to vortices possessing a novel internal structure, which is reflected by the magnetic field associated with these solutions~\cite{multilayered}. In this sense, we believe that the present study, in which we deal with models where the $\rm SU(2)$ symmetry is enhanced to describe  products of several $\rm SU(2)$ factors will also induce interest in planar systems, following the lines of \cite{multilayered}. In the case of vortices, enhancement of the basic $\rm U(1)$ symmetry to a product of $N$ factors, $\rm U(1)\times U(1)\times\cdots\times U(1)$, can be directly connected to the addition of several order parameters in the system. The interest here can be extended to vortices in superconductors described by the Ginzburg-Landau theory and is directly connected with the study of two- and three-component systems that appeared in Refs. \cite{prl2,prl22,prl3}. The subject is also of interest to the case of the multi-component Gross-Pitaevskii equations that describe the mean-field dynamics of spin-1 and spin-2 Bose-Einstein condensates; see, e.g., Ref. \cite{PRep} and references therein for more information on these issues. Moreover, the idea can also be very naturally extended to kinks in the real line, in direct connection with the work \cite{multikink}, in which the $\rm Z_2$ symmetry is enhanced to $\rm Z_2\times Z_2$, and can be used in applications similar to the very recent one, described in \cite{apply}, where geometrically constrained kinks contribute to modify the behavior of fermions, leading to situations of practical interest for the constructions of electronic devices at the nanometric scale. There are other more involved situations, in which the gauge group can appear as a product of the form ${\rm SU(N_1)\times SU(N_2)\times\cdots \times SU(N_p)}$; this is the case, for instance, in the Atiyah, Drinfeld, Hitchin, and Manin construction of multi-instantons with product group gauge theories that arise from D3-branes; see, e.g., \cite{instanton} and references therein.

In this work we will build upon the investigation conducted in~\cite{bimag} to construct multimagnetic structures, which emerge from the superposition of magnetic shells, each of which is the result of spontaneous symmetry breaking in a $\rm{SU(2)}$ subgroup of the full gauge group, under whose transformations our models will be taken to be invariant. Research from the last decade has brought to light a great variety of interesting applications for bimagnetic structures, ranging from engineering to biomedicine~\cite{Estrader, EstraderII, magnetoresistance, Lee}. One particularly interesting application lies in the possibility of using the exchange bias present on multilayers of magnetic material for beating the superparamagnetic limit, which normally constrains the miniaturization of magnetic recording devices~\cite{superpara,superparaII}. Such multilayered or onion-like structures have been investigated in relation to nanoparticles~\cite{Salazar, Catala} and, as multimagnetic structures expand upon the bimagnetic models studied earlier, so we hope that the systems investigated here may be useful to expand on the aforementioned applications. Another interesting property of our models lie in the presence of a Bogomol'nyi-Prasad-Sommerfield (BPS) limit, in which the problem is reduced to the solution of a system of first order differential equations. Solutions of this type are global minima of the energy, and are therefore expected to be stable under small fluctuations.

Magnetic monopoles have also been found in spin ice systems \cite{SI1,SI2} and more recently, the direct observation of static and dynamics of emergent magnetic monopoles in a chiral magnet has been reported in Ref. \cite{PRL20}. They are also object of theoretical and experimental study in heavy-ion collisions and in neutron stars \cite{PRL17}, and also in the international research collaboration MoEDAL running at CERN with the prime goal of searching for the magnetic monopole \cite{MO}. These studies motivate that we investigate new models of magnetic monopoles in high energy physics, hoping to contribute with the construction of models that allow the addition of internal structure, bringing novel properties and/or features at the fundamental level.    

In order to disclose the investigation, we organize the present work as follows: In Sec.~\ref{gen} we introduce the generalized Lagrangian that defines our models, and derive the second order equations that follow from it. We then investigate the conditions under which a Bogomol'nyi bound is attainable, and examine the first order equations which must be satisfied by BPS solutions. We then proceed, in Sec.~\ref{N3} and~\ref{N4}, to solve some examples which engender the sought-after multimagnetic structures, and investigate their properties. Finally, in Sec.~\ref{disc}, we discuss our results and add some perspectives of future works in the subject.

%%%%%%%%%%%%%%%%%%%%%%%%%%%%%%%%%%%%%%%
\section{General Procedure}\label{gen}
We work in four dimensional Minkowski spacetime with metric $(-,+,+,+)$, and consider a class of non-abelian models with the Lagrangian density of the form 
\be\label{lmodel}
\begin{aligned}
\LL &= - \sum_{n=1}^N\frac{P^{(n)}(\{|\phi|\})}{4}F^{(n)a}_{\mu\nu}F^{(n)a\mu\nu} \\
	&\hspace{4mm} -\sum_{n=1}^N\frac{M^{(n)}(\{|\phi|\})}{2} D^{(n)}_\mu \phi^{(n)a} D^{{(n)}\mu} \phi^{{(n)}a}\\
	&\hspace{4mm} - V(\{|\phi|\}),
\end{aligned}
\ee
where $N$ is a positive integer, $(\{|\phi|\})$ stands for the set $(|\phi^{(1)}|,|\phi^{(2)}|,...,|\phi^{(N)}|)$, $D^{(k)}_\mu \phi^{(k)a} =\partial_\mu\phi^{(k)a} + g^{(k)}\veps^{abc}A^{(k)b}_\mu\phi^{(k)c}$ is the action of covariant derivative in the $k$-th sector with coupling $g^{(k)}$ and $F^{(k)a}_{\mu\nu} = \partial_\mu A^{(k)a}_\nu - \partial_\nu A^{(k)a}_\mu + g^{(k)}\veps^{abc}A^{(k)b}_\mu A^{(k)c}_\nu$ is the $k$-th field strength. The potential is denoted by $ V(\{|\phi|\})$. 

This Lagrangian is invariant under the action of the product of $N$ $\rm{SU}_{(k)}(2)$ groups, in the form  $\rm{SU}_{(1)}(2)\times ...\times \rm{SU}_{(N)}(2)$. Each scalar field is a $\rm{SU(2)}$-valued quantity, given by $\phi^{(n)}=\phi^{(n)a}T^{a}$, where $T^a=-i\sigma^a/2$ are the generators of SU(2), given in terms of the three Pauli matrices $\sigma^a$. Each $\phi^{(n)}$ is charged only in the respective $\rm{SU}_{(n)}(2)$ subgroup, and is thus coupled to a gauge field $A^{(n)}_\mu= A^{(n)a}_\mu T^a$. The condition $\langle T^a, T^b\rangle = -2{\rm{Tr}}(T^aT^b)=\delta^{ab}$ may be used to define an inner product in SU(2), under which one obtains the norm $|\phi^{(n)}|=\sqrt{\phi^{(n)a}\phi^{(n)a}}$.  Throughout this work, the Einstein summation convention is always implied for Lorentz and SU(2) internal indices, but not for the remaining ones, which simply label the $\rm{SU}(2)$ subgroups and are always enclosed in parentheses.

The functions $P^{(n)}$ and $M^{(n)}$ are nonnegative and may, in principle, depend on all the scalar fields $\{|\phi|\}$, respecting the gauge symmetry, but we shall shortly restrict their forms, aimed to implement some possibilities capable of giving interesting stable minimum energy configurations. We generically require that these functions take finite values when the field configurations approach the vacuum manifold, which will be topologically non trivial, in order to allow for finite energy topological solutions. {\color{red} }The solutions can be classified according to a set of $N$ integers, which, in analogy to the standard monopole, we may relate to quantized \qt{magnetic} charges emerging from the unbroken $\rm{U(1)}$ symmetry from each $\rm{SU(2)}$ submanifold.

The field equations derived from ~\eqref{lmodel} for the $k$-th sector are
\bes\label{geom}
	\begin{align}
		&D^{(k)}_\mu\left(M^{(k)} D^{(k)\mu}  \phi^{(k)a}\right) =\frac{1}{4}\sum_{n=1}^N\pd{P^{(n)}}{\phi^{(k)a}}F^{(n)b}_{\mu\nu}F^{(n)b\mu\nu} \nonumber\\
		& +\frac{1}{2}\sum_{n=1}^N\pd{M^{(n)}}{\phi^{(k)a}} D^{(n)}_\mu \phi^{(n)b} D^{{(n)}\mu} \phi^{{(n)}b} + \pd {V}{\phi^{(k)a}},\\ 
	& D^{(k)}_\mu\left(P^{(k)}F^{(k)a\mu\nu}\right) =  g^{(k)}M^{(k)}\,\veps^{abc}\phi^{(k)b} D^{(k)\nu} \phi^{(k)c},
	\end{align}
	\ees
where the covariant derivatives also act in the following form $D^{(n)}_\mu F^{(n)a\mu\nu} = \partial_\mu F^{(n)a\mu\nu} + g^{(n)}\veps^{abc}A^{(n)b}_\mu F^{(n)c\mu\nu}$.

Let us now consider static solutions, which may be obtained as time independent fields which obey equations of motion in a gauge such that $A_{0}^{(n)}=0$ (leading to the non Abelian analogues of Gauss' law to be trivially solved). The energy functional in this case is given by
\begin{equation}\label{Energyfunc}
\begin{split}
E= \sum_{n=1}^{N}&\int d^3x\left[\frac{1}{2}M^{(n)}|D_k^{(n)}\phi^{(n)}|^2 + \frac{1}{2}P^{(n)}|B_k^{(n)}|^2\right. \\
& \left. + V(\{|\phi|\})\right],
\end{split}
\end{equation}
where $B_k^{(n)}\equiv\varepsilon_{ijk}F^{ij(n)}/2$ and the previously defined SU(2) norm has been used. We may now follow~\cite{bogo} to find a first order framework. This can be achieved by setting  $M^{(n)}=1/P^{(n)}$ for each $n$ and taking the limit $V(\{|\phi|\})\to 0$ in~\eqref{Energyfunc}. This leads to
\begin{equation}\label{EnergyfuncII}
\begin{split}
E=& \sum_{n=1}^{N}\int d^3x\; P^{(n)}\;\left| B_{k}^{(n)} \pm  \frac{D_k^{(n)}\phi^{(n)}}{P^{(n)}}\right|^2 \\
&+ 4\pi\sum_{n=1}^N \frac{v^{(n)}}{g^{(n)}}|Q_t^{(n)}|,
\end{split}
\end{equation}
where $Q_t^{(n)}\in \mathbb{Z}$  is the topological charge associated with the $(\phi^{(n)}, A_{\mu}^{(n)})$ solutions, defined in the same way as in the standard $\rm{SU}(2)$ Yang-Mills-Higgs theory~\cite{manton}. The second term in the right hand side is recognized as the summation of the fluxes arising from the magnetic fields in each sector. The observation that the first integral in~\eqref{EnergyfuncII} is non-negative leads to the Bogomol'nyi bound
\begin{equation} \label{Bound}
E \geq 4\pi\sum_{n=1}^N \frac{v^{(n)}}{g^{(n)}}|Q_t^{(n)}|.
\end{equation}
The equality in~\eqref{Bound} is attained if and only if the first order equations

\begin{equation}\label{BPSgen}
B_k^{(n)}=\mp \frac{D_k^{(n)}\phi^{(n)}}{P^{(n)}}
\end{equation}
are satisfied for each $n$. By direct derivation it is straightforward to show that solutions of~\eqref{BPSgen} always solve the second order equations. Deviations from the static case give rise to quadratic terms in $F_{j0}^{(n)}$ and $\pb{0}\phi^{(n)}$ which must be added to the integral in~\eqref{Energyfunc}. Such terms may only increase the energy, and thus configurations that solve~\eqref{BPSgen} are global minima of this functional.

We shall henceforth concern ourselves with spherically symmetric static solutions. Since we are choosing  $A_{0}^{(n)}=0$, as is well known, these restrictions allow us to write the fields in the form~\cite{thooft,polyakov}
\bes\label{ansatz}
\begin{align}
\phi^{(n)a} &= \frac{x_a}{r} H^{(n)}(r), \\  A_i^{(n)a} &= \veps_{aib}\frac{x_b}{g^{(n)}r^2}(1-K^{(n)}(r)),
\end{align}
\ees
subject to the conditions
\be\label{bcond}
\begin{aligned}
H^{(n)}(0)&=0, & K^{(n)}(0)&=1,\\
H^{(n)}(\infty) &\to \pm v^{(n)}, & K^{(n)}(\infty) &\to 0.\\
\end{aligned}
\ee

These requirements ensure that the ensuing energy is finite, while also enforcing symmetry breaking and thus resulting in topological solutions. The form of this Ansatz implies $|Q_{t}^{(n)}|=1$ for all $n$, so that the magnetic charge in each subsystem achieves the smallest non-zero value allowed by the quantization condition. In the standard magnetic monopole the vacuum manifold is the 2-sphere $S^2$, but in the present model it is the product of $N$ 2-sphere, being $S^2\times S^2\times \cdots \times S^2$. As one knows, the second homotopy group of $S^2$ is the group of the integers, $Z$; however, we also know that the second homotopy group of the product of $N$ 2-sphere $S^2$ is the product of $N$ second homotopy group of the 2-sphere $S^2$, so we have $Z\times Z\times \cdots \times Z$, that is, a set of $N$ integers. In this sense, the classification by a set of $N$ integers exhausts the finite energy solutions, including in particular the spherically symmetric solutions that we will describe below. Under the above assumptions, the second order equations take the form
\bes\label{geomansatz}
\begin{equation}
\begin{split}
&\frac{1}{r^2}\left(r^2M^{(n)} {H^{(n)}}^\prime\right)^\prime - \frac{2M^{(n)} H^{(n)} {K^{(n)}}^2}{r^2} \\
&-\frac{1}{2}\sum_{k=1}^N\pd{P^{(k)}}{H^{(n)}} \left(\frac{2{{K^{(k)}}^\prime}^2}{{g^{(k)}}^2r^2} - \frac{(1-{K^{(k)}}^2)^2}{{g^{(k)}}^2r^4}\right)\\
&-\frac{1}{2}\sum_{k=1}^N\pd{M^{(k)}}{H^{(n)}}\left({{H^{(k)}}^\prime}^2 + \frac{2{H^{(k)}}^2{K^{(k)}}^2}{r^2}\right) \\ &- \pd{V}{H^{(n)}}=0,
\end{split}
\end{equation}
and
\begin{equation}
\begin{split}
& r^2\left(P^{(n)} {K^{(n)}}^\prime\right)^\prime -  K^{(n)}\left\{\left(M^{(n)} {g^{(n)}}^2r^2{H^{(n)}}^2\right)\right. \\
& \left. + \left(P^{(n)}\,(1-{K^{(n)}}^2)\right) \right\}=0,
\end{split}
\end{equation}
\ees
where the  prime denotes differentiation with respect to the radial coordinate. These equations constitute a set of $2N$ coupled and nonlinear ordinary differential equations. This system is very hard to solve, even when we choose simple couplings among the $N$ sectors with ${\rm SU(2)}$ symmetry to complete the model. For this reason, let us substitute the Ansatz into~\eqref{BPSgen} and take the same assumptions used in that derivation to find the first order equations in the presence of spherical symmetry. For each $n$, there result two independent equations, obtained by projecting~\eqref{BPSgen} into directions parallel and orthogonal to $\phi^{(n)}$, so that the energy is minimized by configurations that solve the $N$ pairs of first order equations

\bes\label{foh}
\bal
&{H^{(n)}}^\prime =\pm P^{(n)} \frac{1-{K^{(n)}}^2}{g^{(n)}r^2},\\
&{K^{(n)}}^\prime =\mp g^{(n)}\frac{H^{(n)}K^{(n)}}{P^{(n)}},
\eal
\ees
for $n=1,...,N$. These configurations saturate the Bogomol'nyi bound, with energy
\be \label{Energy}
E = 4\pi\sum_{k=1}^N \frac{v^{(k)}}{g^{(k)}},
\ee
so that each $\rm{SU(2)}$ subsystem contributes with an additive factor $4\pi v^{(k)}/g^{(k)}$ to the total energy. This, of course, is a reflection of the fact that these solutions have unity topological charges in each sector. We also note that the energy does not depend on any of the $P^{(n)}$, so the Bogomol'nyi bound only depends on the asymptotic values of the fields, which always approach the trivial vacuum solutions when the boundary conditions \eqref{bcond} are satisfied.

We may choose the functions $P^{(n)}$ in a way that makes it easier to solve the first order equations while still keeping the subsystems coupled in a non-trivial way. In order to achieve this, we can consider several distinct possibilities, some of them will be considered below to illustrate the general situation. Before doing this, however, let us notice that the energy density for solutions of the first order equations can be written in the form
\begin{equation}
\rho(r)=\sum_{n=1}^N\rho_n(r),
\end{equation}
where
\begin{equation}
\begin{split}
	\rho_n(r)=& \frac{2P^{(n)}{K^{(n)}}^{\prime 2}}{(g^{(n)}r)^2} + \frac{{H^{(n)}}^{\prime 2}}{P^{(n)}} ,\label{rhoa}\\
	=&\frac{P^{(n)}(1-K^{(n)2})^2}{(g^{(n)}r^2)^2} + \frac{2(H^{(n)}K^{(n)})^2}{r^2P^{(n)}},
\end{split}
\end{equation}
which represents the energy density for the $n$-th sector.
In this sense, since we are interested in constructing multimagnetic monopoles, which are localized structures composed of several layers with distinct magnetic profiles, we then concentrate on the case
\begin{equation}\label{P}
    P^{(n)}(\{|\phi|\}) =\prod_{k=1}^N \left(\frac{1}{|\phi^{(k)}|}\right)^{\alpha_k^n},
\end{equation}
where the real and non-negative quantities $\alpha_k^n$ identify the set of parameters that adds interaction among the $N$ subsystems. Each particular choice of the set $\{\alpha^n_k\}$ describe a specific model, with the understanding that the zeroes of $\{\alpha^n_k\}$ remove the dependence of $P^{(n)}$ on the $k$-th scalar field. As we will show below, the specific values of the non vanishing parameters contribute to identify the maximum and width of the energy density associated to the corresponding magnetic shell in the multimagnetic structure.

\section{MODELS WITH N=3}\label{N3}
Let us first illustrate the aforementioned procedure with a few examples, in which the Lagrangian is of the form \eqref{lmodel}, with gauge group $\rm{SU}(2)\times \rm{SU}(2)\times \rm{SU}(2)$. For simplicity, let us take $g^{(1)}=g^{(2)}=g^{(3)}=1$ and $v^{(1)}=v^{(2)}=v^{(3)}=1$ in our calculations and, since the upper and lower signs in equations \eqref{foh} are related by changing $H\to -H$, we then only consider the upper signs in the investigation that follows.

\subsection{First Model}\label{ex1}
As a first example, we shall consider a model with a standard monopole core, meaning $P^{(1)}=1$. According to the form suggested in Eq. \eqref{P}, we are sending $\alpha_1^1$, $\alpha_2^1$ and $\alpha_3^1$ to zero. The $k=1$ equations in~\eqref{foh} are thus
\begin{subequations}\label{BPS1}
\begin{align}
H^{(1)'}&=\frac{1- K^{(1)}}{r^2}, \\
K^{(1)'}&=-H^{(1)}K^{(1)},
\end{align}
\end{subequations}
which are simply the Bogomol'nyi equations for the 't Hooft-Polyakov monopole, for which an analytic solution has long been known~\cite{ps}:
\begin{align}\label{standard}
H^{(1)}(r)=\coth(r) - \frac{1}{r}, && K^{(1)}(r)=r\;\rm{csch}(r).
\end{align} 
The energy density for this solution can be found from~\eqref{rhoa} to be
\begin{equation}\label{rho1}
\rho_1(r)=\frac{[r^2 - \sinh^2(r)]^2}{r^4\sinh^4(r)} +\frac{2(r\coth(r)-1)^2}{r^2\sinh^2(r)}.
\end{equation}

We may now proceed as in~\cite{bimag} and introduce a shell structure by the choices $P^{(2)}=|\phi^{(1)}|^{-\alpha}$ and $P^{(3)}=|\phi^{(2)}|^{-\beta}$, where $\alpha=\alpha^2_1$ and $\beta=\alpha^3_2$ are now positive integers, with all the other parameters vanishing. With these choices, we are led to the equations
\bes\label{3b}
\begin{align}
&H^{(2)'}=\left(\frac{1}{r\coth(r)- 1}\right)^{\alpha}\frac{1-{K^{(2)}}^2}{r^{2-\alpha}},\\
&K^{(2)'}=-H^{(2)}K^{(2)}\left(\coth(r) - \frac{1}{r}\right)^{\alpha}.
\end{align}
\ees
and
\bes\label{3c}
\begin{align}\label{BPS3}
&H^{(3)'}=\left(\frac{1}{H^{(2)}}\right)^{\beta}\frac{1-{K^{(3)}}^2}{r^{2}},\\
&K^{(3)'}=-H^{(3)}K^{(3)}\left(H^{(2)}\right)^{\beta}.
\end{align}
\ees
Note that we have inserted the solution~\eqref{standard} for $H^{(1)}(r)$ in equations~\eqref{3b}, resulting in a system of two first order equations which can now be solved for the two unknown functions $H^{(2)}$ and $K^{(2)}$. We have not been able to obtain $H^{(2)}$ and $K^{(2)}$ in closed form, but the problem is amenable to numerical analysis.

\begin{figure}[ht]
	\centering
	\includegraphics[scale=0.21]{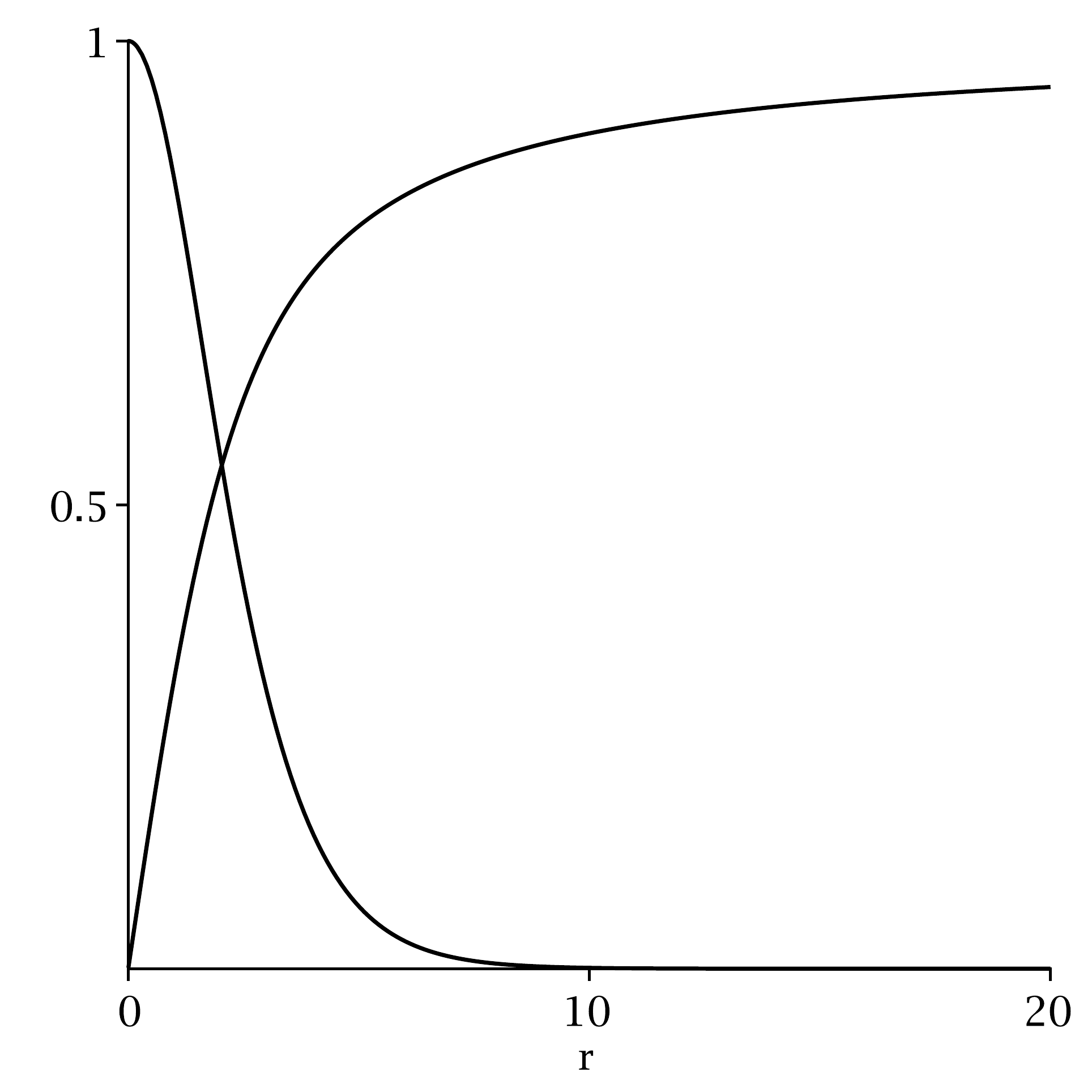}
	\includegraphics[scale=0.21]{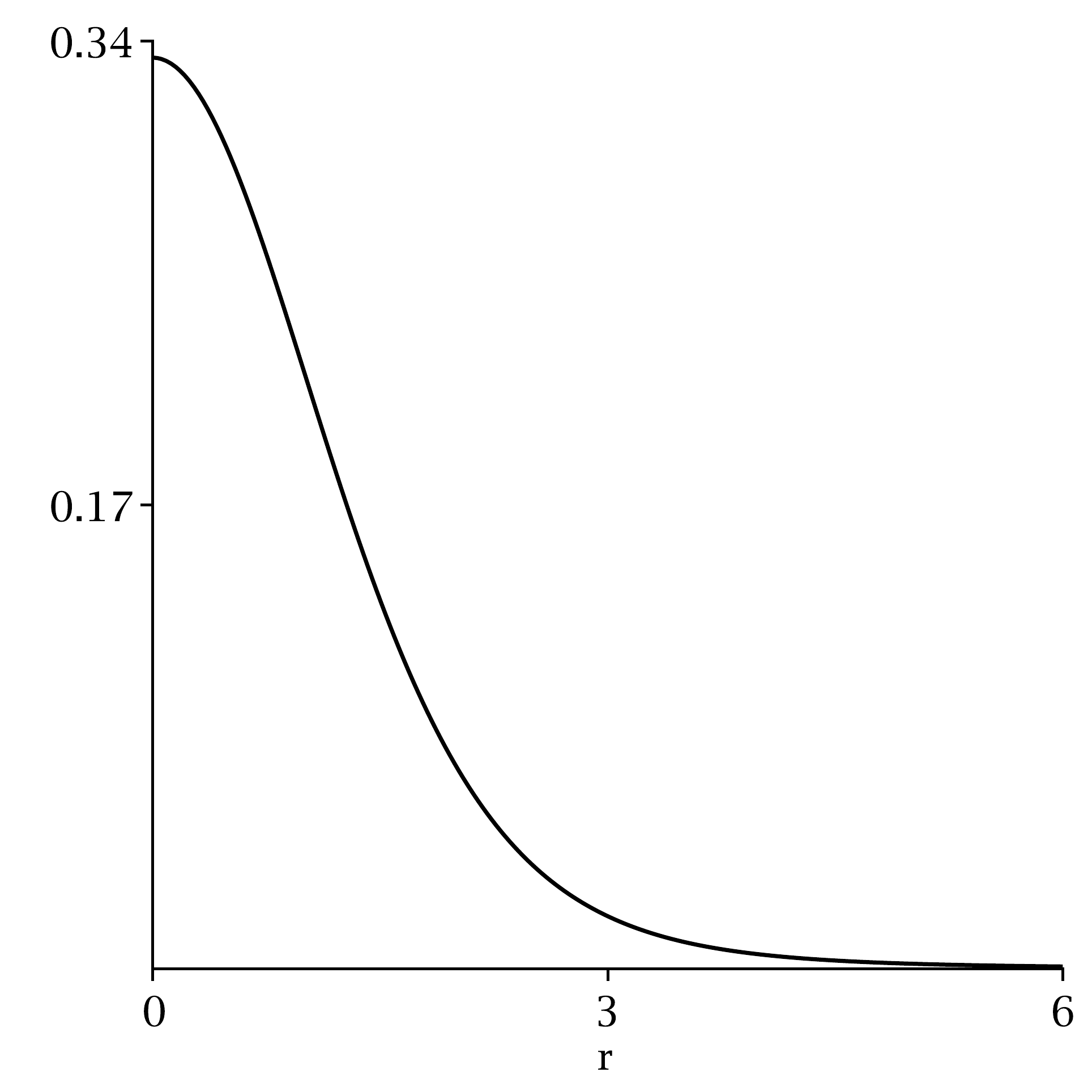}
	\caption{In the left panel, we show the solutions $H^{(1)}(r)$ (ascending line) and $K^{(1)}(r)$ (descending line) of~\eqref{BPS1}. Their analytical form is given by~\eqref{standard}. In the right panel, we show the contribution $\rho_1$ to the energy density, given by~\eqref{rho1}.}
	\label{profile1}
\end{figure}

\begin{figure}[ht]
	\centering
\includegraphics[scale=0.21]{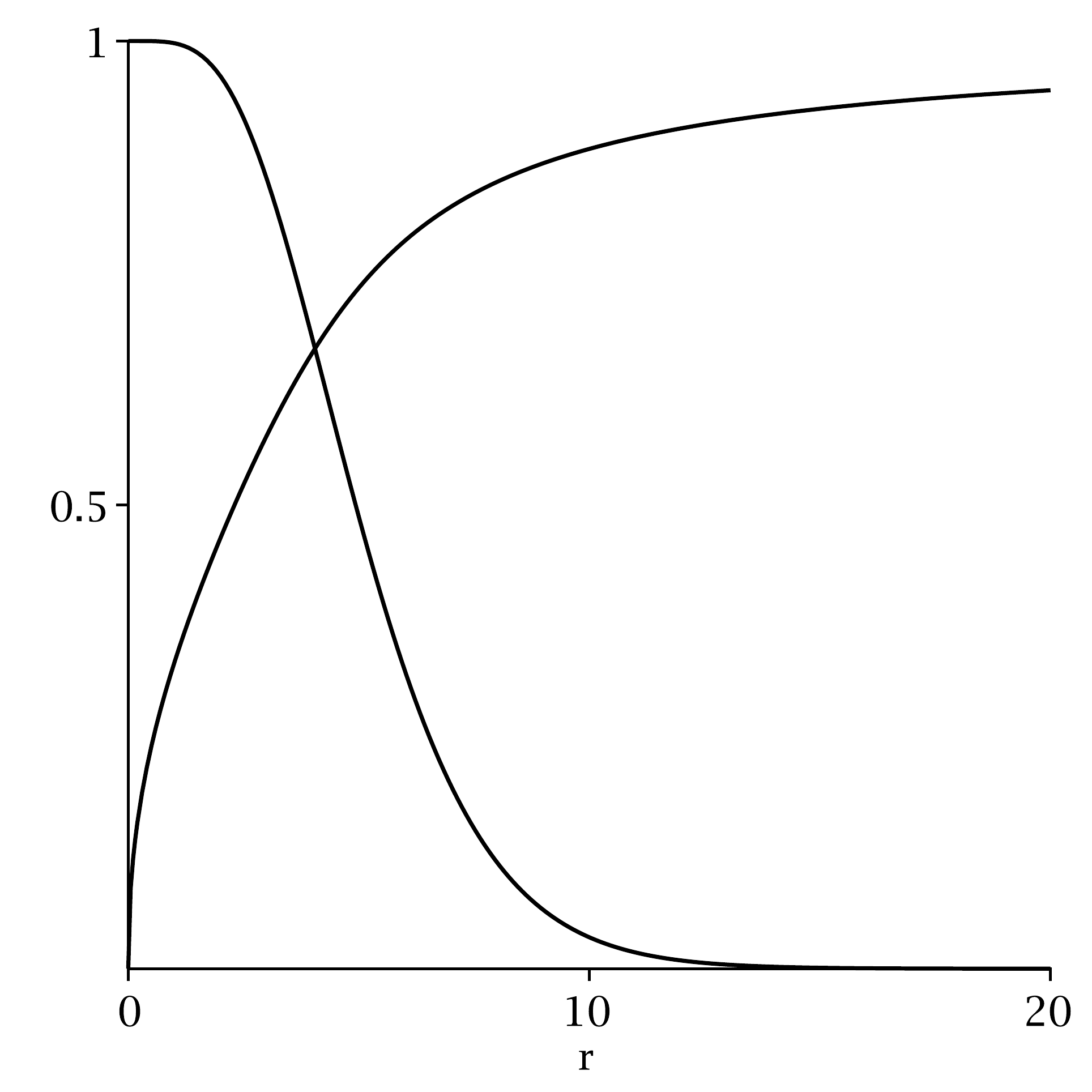}
\includegraphics[scale=0.21]{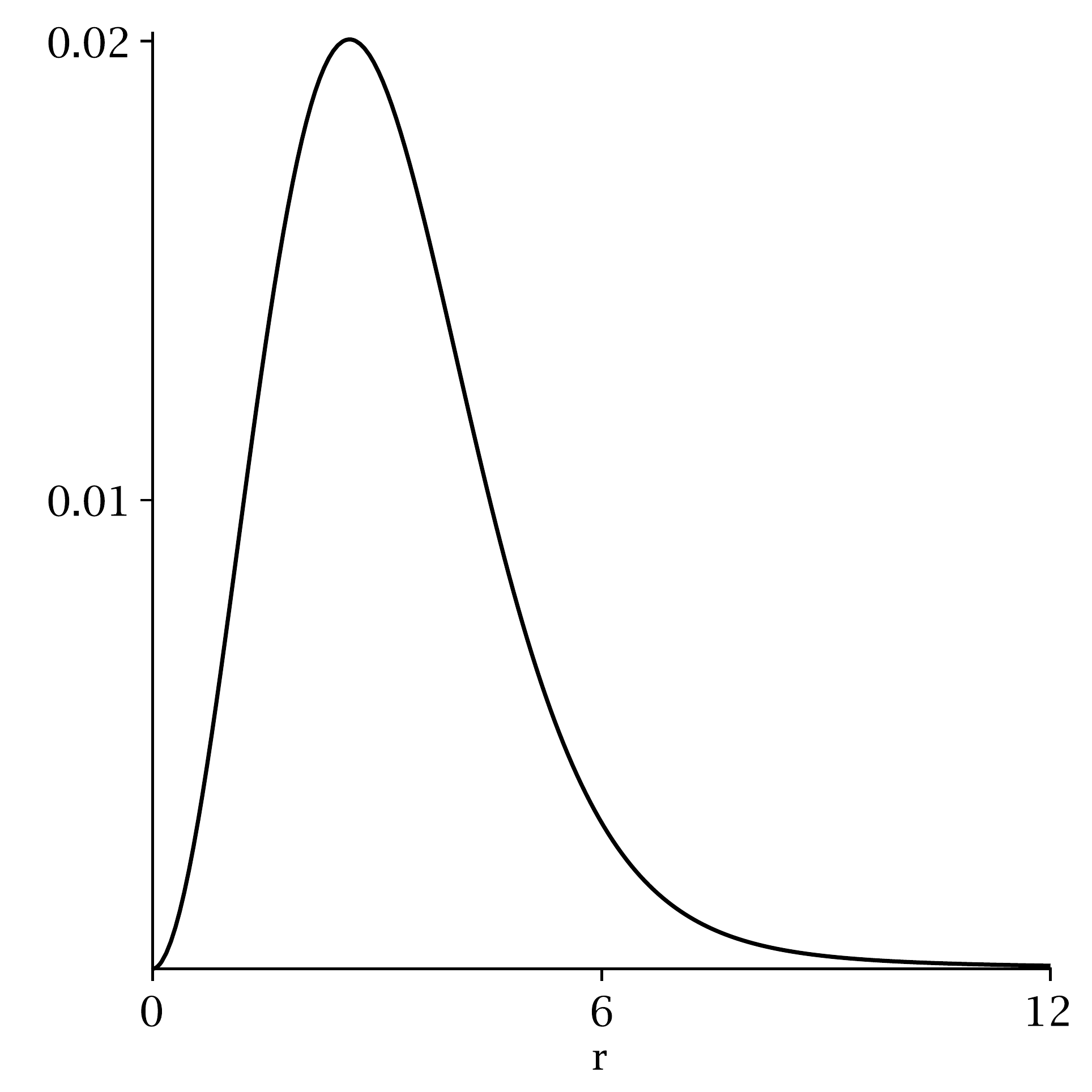}
	\caption{In the left panel, we show the solutions $H^{(2)}(r)$ (ascending line) and $K^{(2)}(r)$ (descending line) of~\eqref{3b} for $\alpha=3$. In the right panel, we show the contribution $\rho_2$, calculated from~\eqref{rhoa}, relative to this sector.}
	\label{profile2}
\end{figure}

\begin{figure}[ht]
	\centering
\includegraphics[scale=0.21]{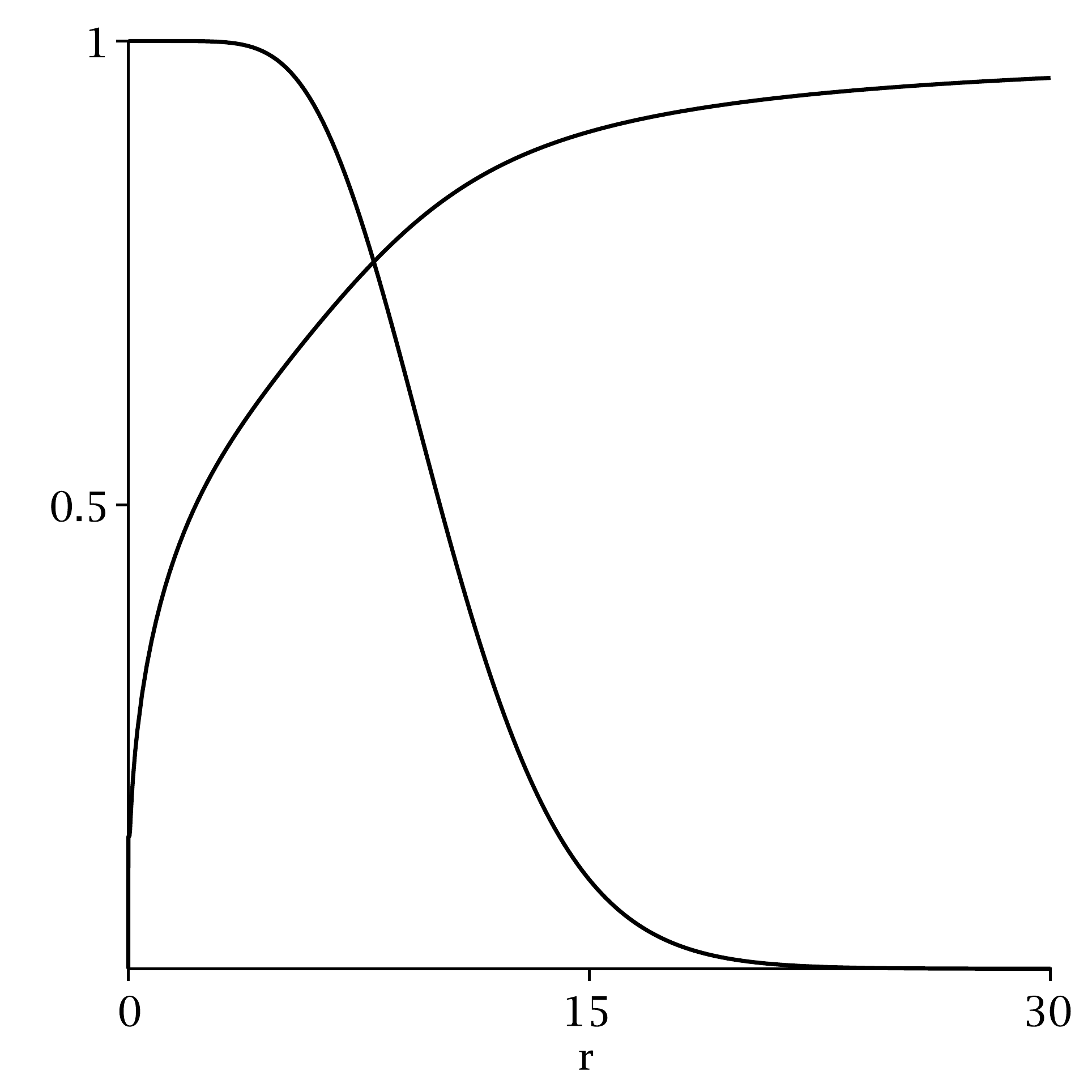}
\includegraphics[scale=0.21]{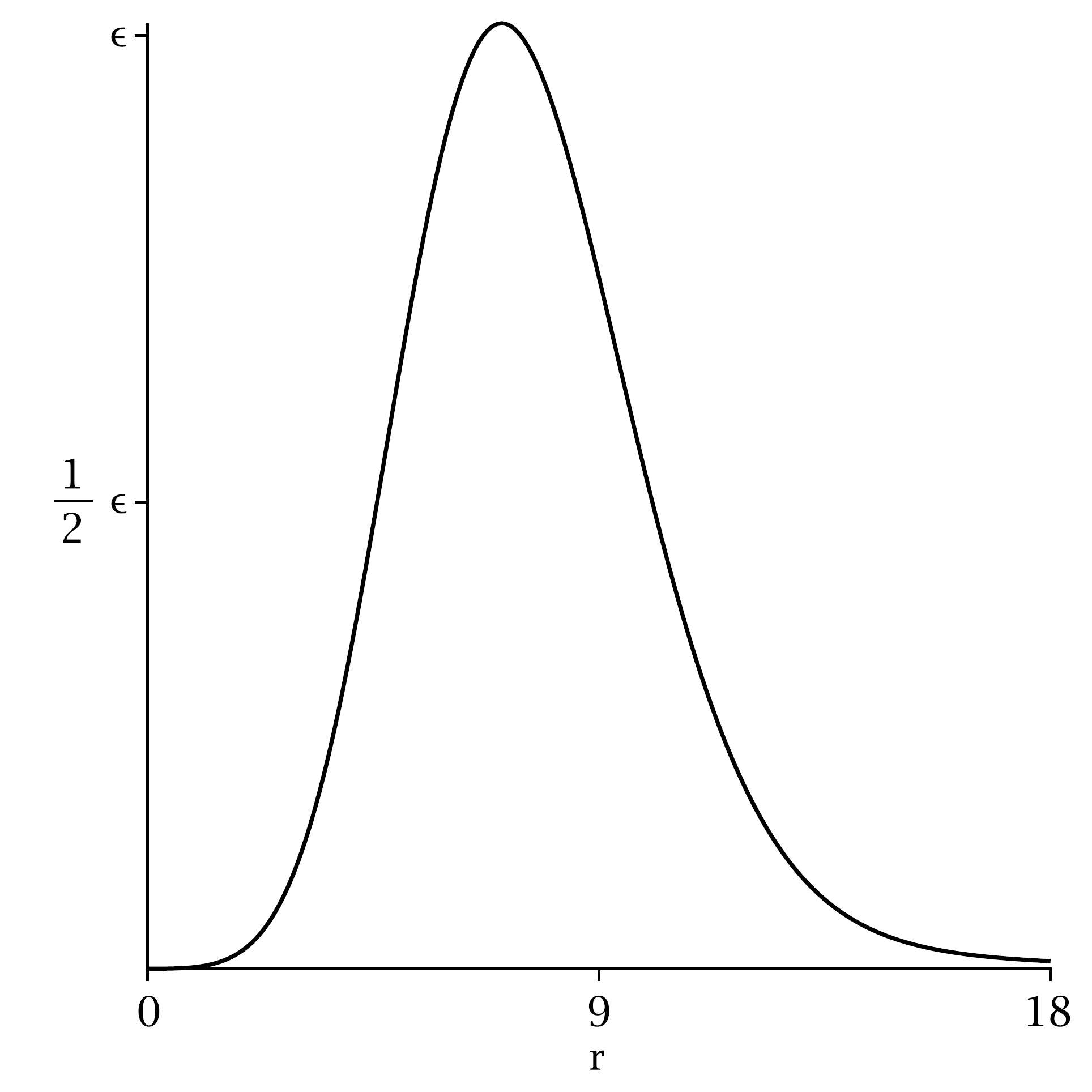}
	\caption{In the left panel, we show the solutions $H^{(3)}(r)$ (ascending line) and $K^{(3)}(r)$ (descending line) of~\eqref{3c} for $\alpha=3$ and $\beta=10$. In the right panel, we show the contribution $\rho_3$, calculated from~\eqref{rhoa}, relative to this sector. Here, $\epsilon=0.0024$.}
	\label{profile3}
\end{figure}

Once the full equations are solved numerically, we may plug the result $H^{(2)}$ in~\eqref{3c} in order to solve these equations. We have done this for the case $\alpha=3$ and $\beta=10$, and present the results in Figs.~\ref{profile1},~\ref{profile2} and~\ref{profile3}. One sees that in all cases these functions present a behavior that is qualitatively similar to that of the standard 't Hooft-Polyakov Monopole. The contributions $\rho_1$, $\rho_2$ and $\rho_3$ to the energy density can be calculated from~\eqref{rhoa}. Their behavior are represented in Figs.~\ref{profile1},~\ref{profile2} and~\ref{profile3}. The last two contributions are strikingly different from the standard monopole solution (which has energy density equal to what we here have called $\rho_1$), presenting a maximum for $r\neq 0$.


\begin{figure}[ht]
	\centering
	\includegraphics[scale=0.7]{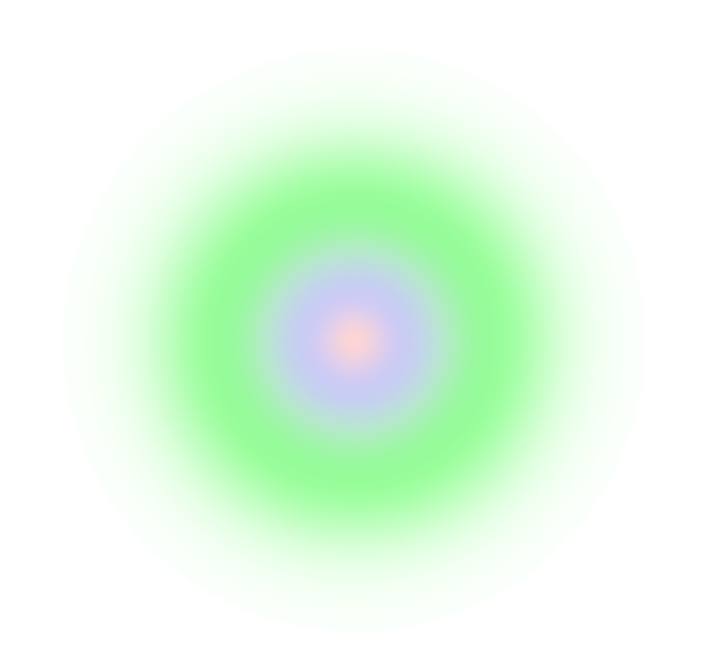}
	\caption{Planar section, passing through the center, of the energy density, in which the contributions $\rho_1$ (red), $\rho_2$ (blue) and $\rho_3$ (green) calculated in subsection~\ref{ex1} are shown, giving rise to a multimagnetic structure. We depict the case $\alpha=3$ and $\beta=10$.}
	\label{three}
\end{figure}

To help us understand the above results, we depict in Fig.~\ref{three} a planar section of the energy density, in which the contributions $\rho_1$, $\rho_2$ and $\rho_3$ have been plotted simultaneously, with the colors red, blue and green, respectively. We see how a novel structure, comprised by a standard monopole core and two shells, arises from the superposition of the different subsystems. The thickness of the shells and the holes associated to them are controlled by the choice of the parameters $\alpha$ and $\beta$, and the maxima of $\rho_1(r)$, $\rho_2(r)$ and $\rho_3(r)$ are attained at $r=0$, $r\approx 2.6$ and $r\approx7.1$, respectively.

The total energy could be calculated by direct integration of $\rho=\rho_1 + \rho_2 + \rho_3$, but these solutions solve the first order equations, so we may simply apply~\eqref{Energy} to find $E=12\pi$.

%%%%%%%%%%%%%%%%%%%%%%%%%%%%%%%%
\subsection{Second Model}\label{second}

\begin{figure}[ht]
	\centering
	\includegraphics[scale=0.21]{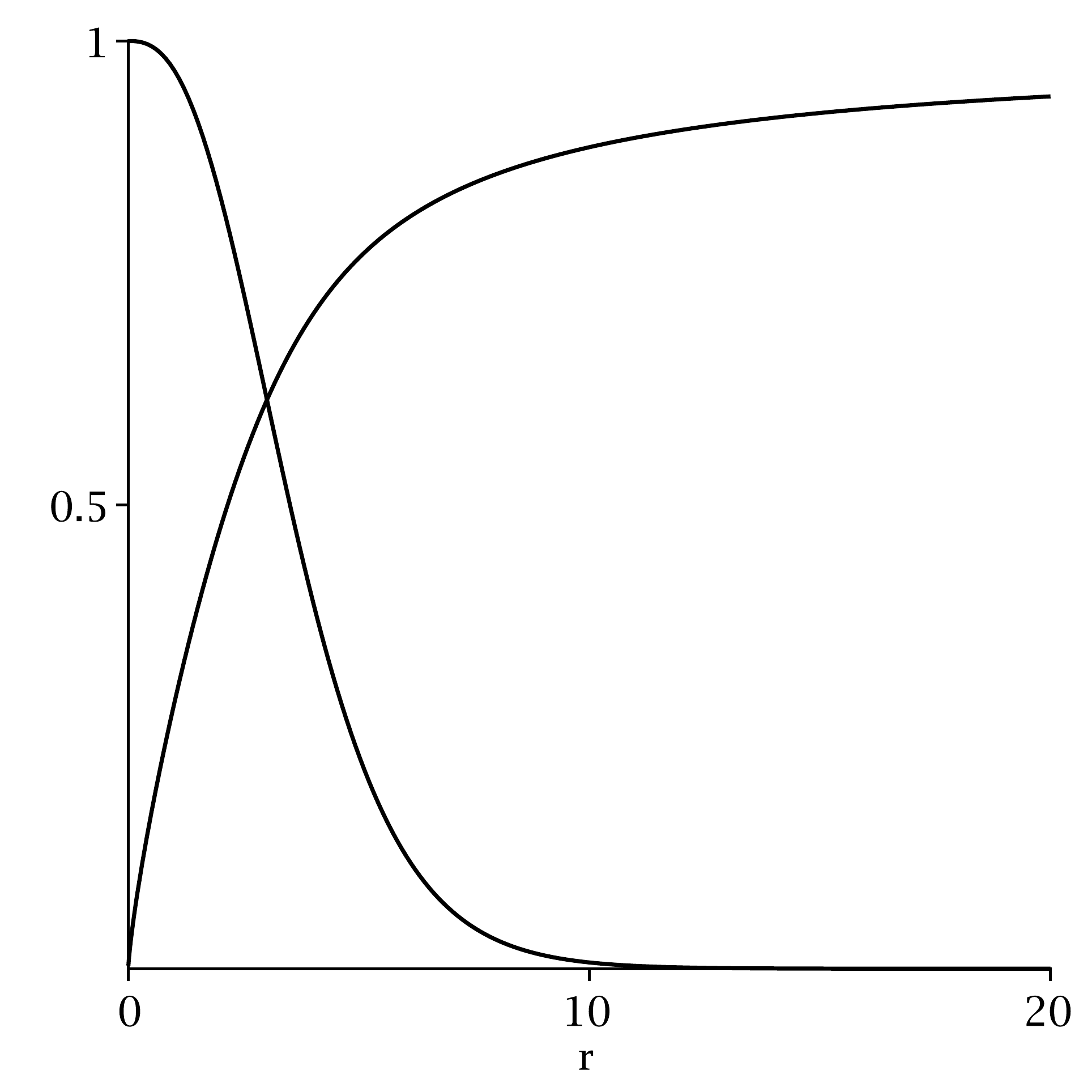}
	\includegraphics[scale=0.21]{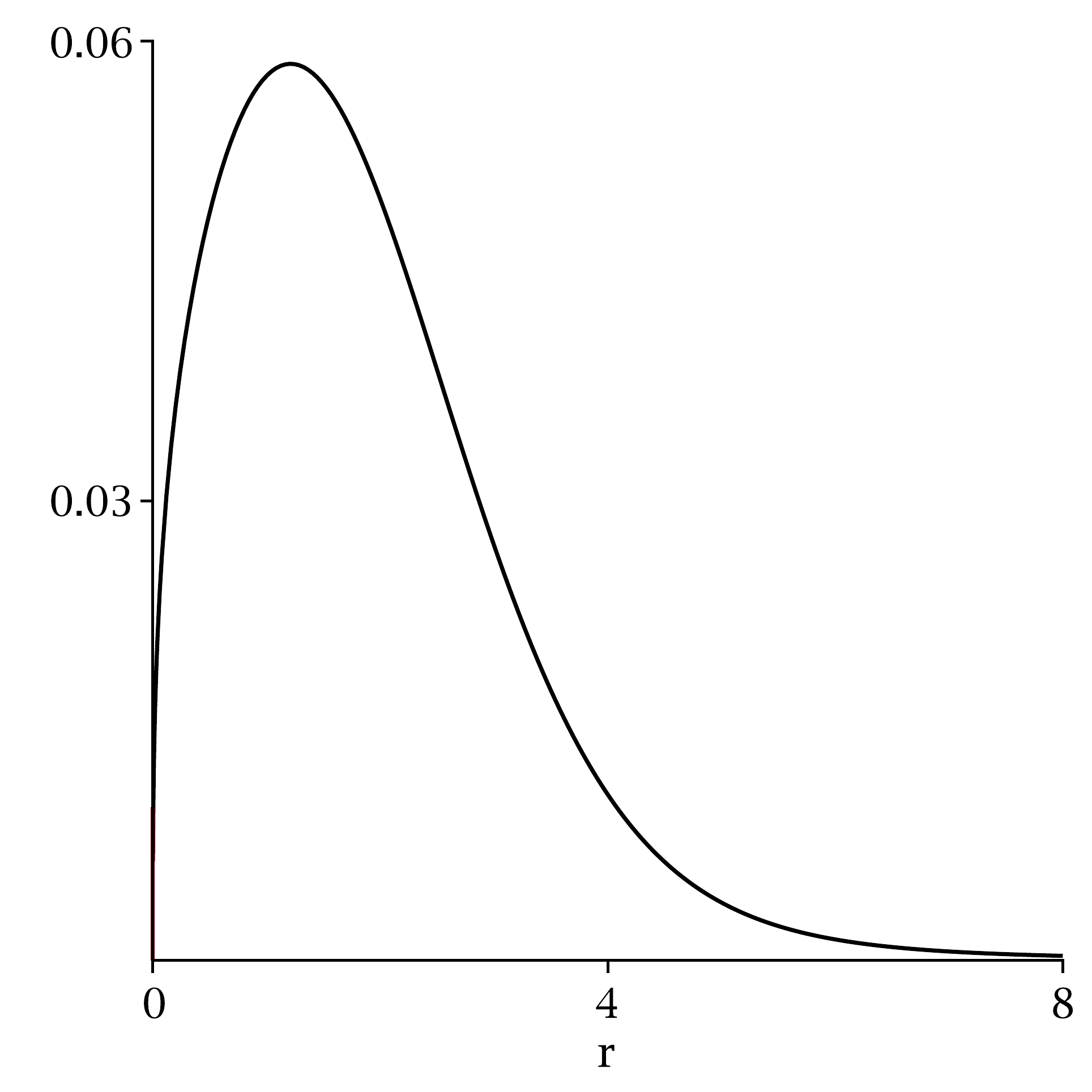}
	\caption{In the left panel, we show the solutions $H^{(1)}(r)$ (ascending line) and $K^{(1)}(r)$ (descending line) of ~\eqref{3d} for $\alpha=1$. In the right panel, we show the contribution $\rho_1$, calculated from~\eqref{rhoa}, relative to this sector.}
	\label{profile4}
	%\end{center}
\end{figure}

\begin{figure}[ht]
	\centering
		\includegraphics[scale=0.21]{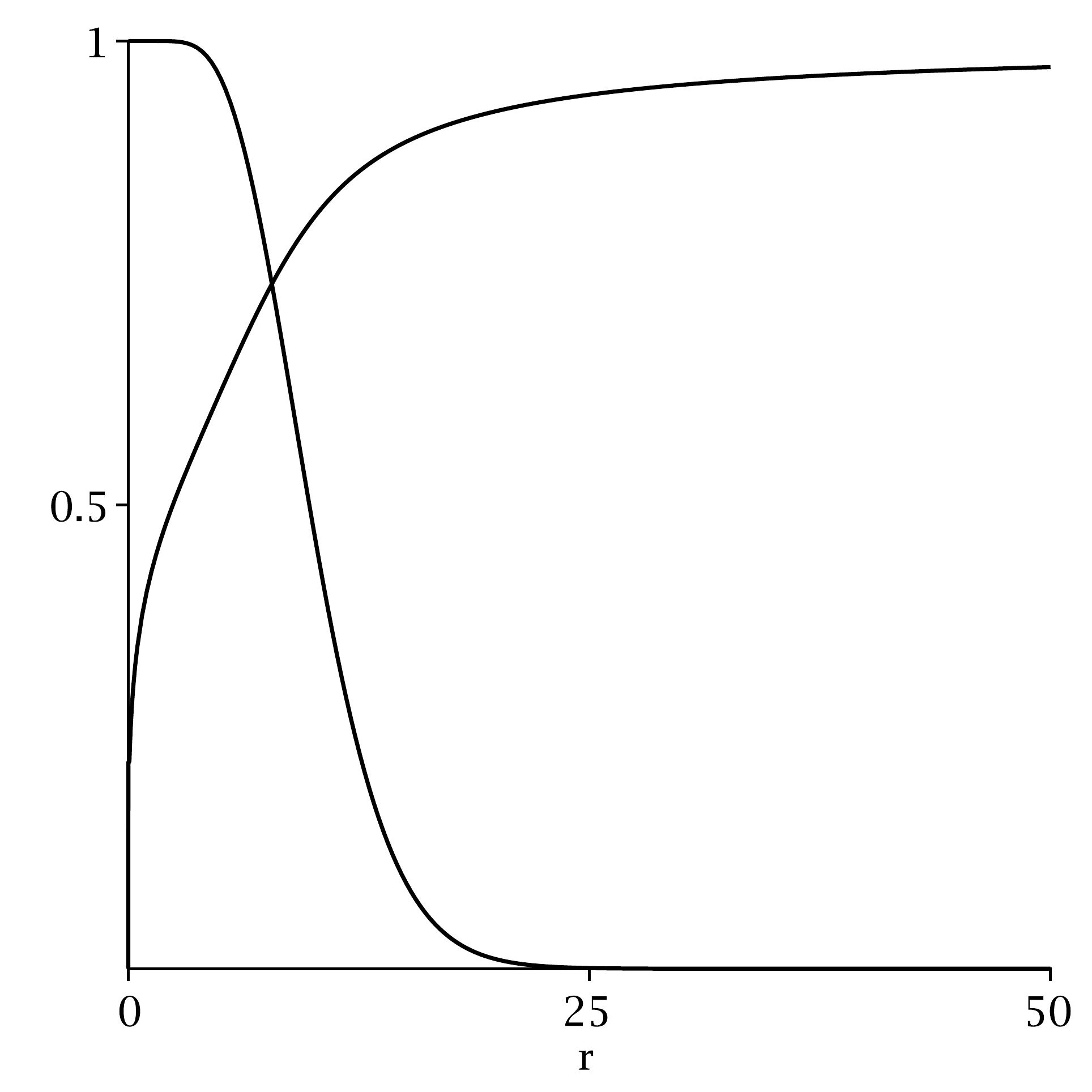}
		\includegraphics[scale=0.21]{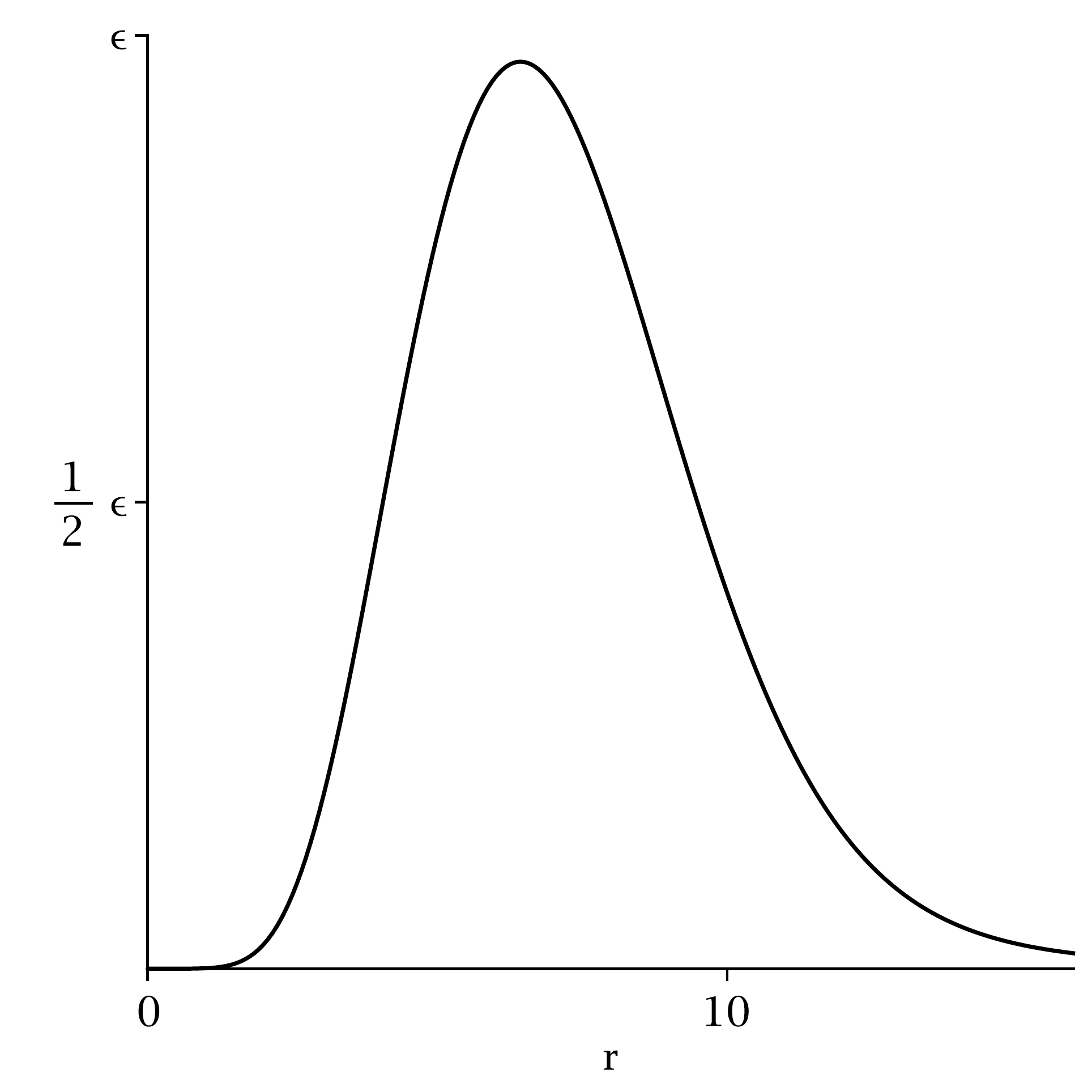}
	\caption{In the left panel, we show the solutions $H^{(2)}(r)$ (ascending line) and $K^{(2)}(r)$ (descending line) of ~\eqref{3e} for $\alpha=1,$ and $\beta=10$. In the right panel, we show the contribution $\rho_2$, calculated from~\eqref{rhoa}, relative to this sector. Here  we have $\epsilon=0.0027$.}
	\label{profile5}
	%\end{center}
\end{figure}

\begin{figure}[ht]
	\centering
\includegraphics[scale=0.21]{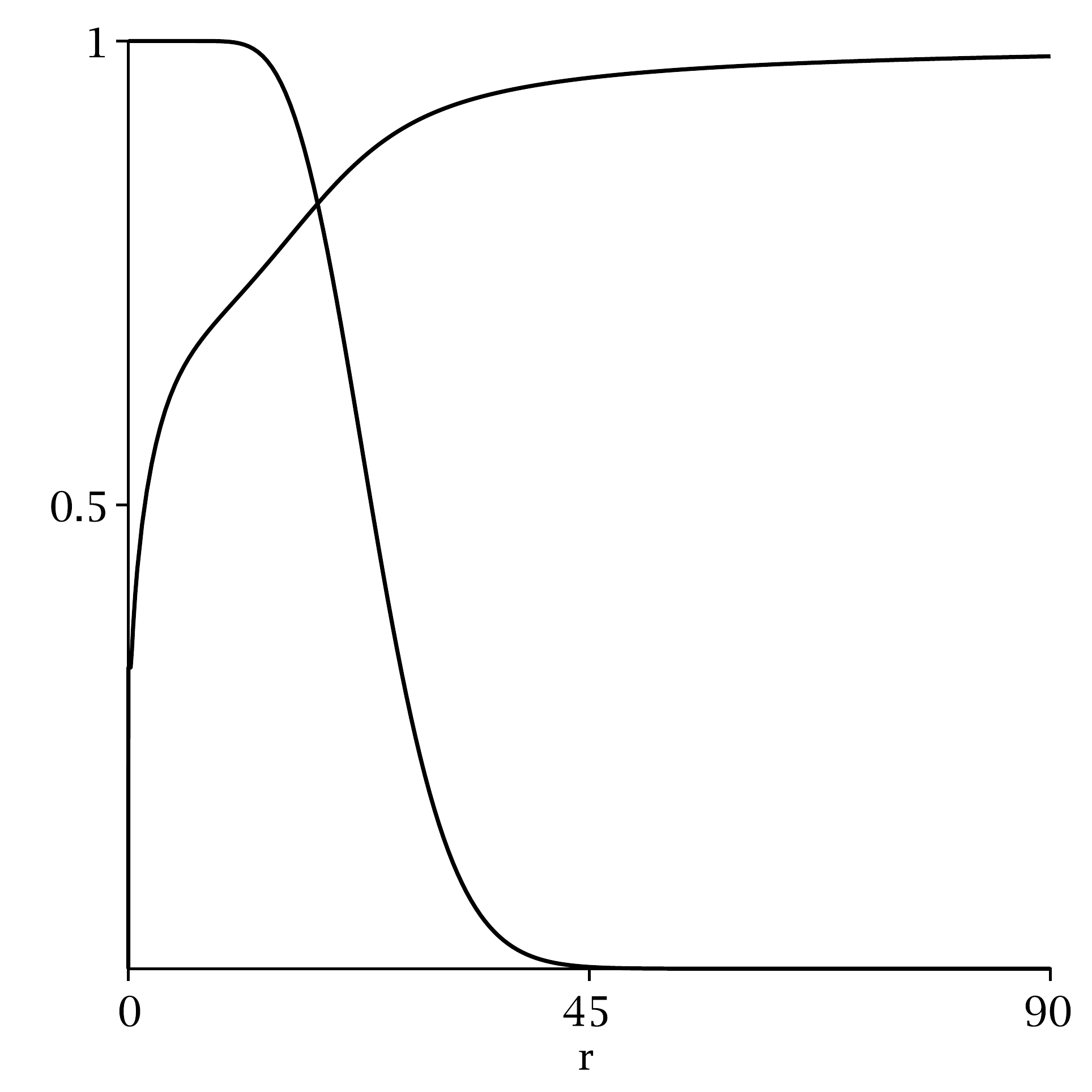}
\includegraphics[scale=0.21]{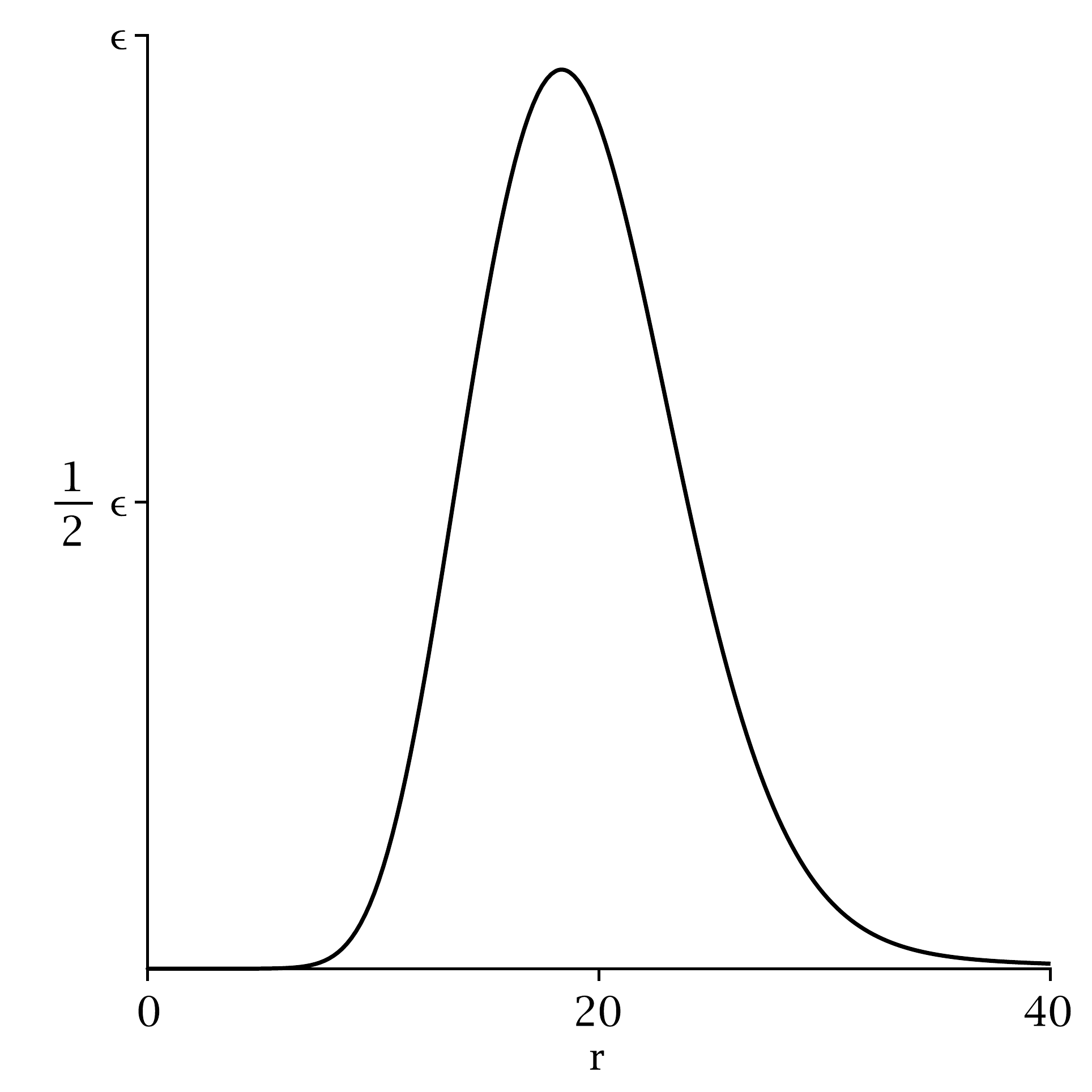}
	\caption{In the left panel, we show the solutions $H^{(3)}(r)$ (ascending line) and $K^{(3)}(r)$ (descending line) of ~\eqref{3f} for $\alpha=1, \beta=10$ and $\gamma=30$. In the right panel, we show the contribution $\rho_3$, calculated from~\eqref{rhoa}, relative to this sector. Here, $\epsilon=0.00022$.}
	\label{profile6}
	%\end{center}
\end{figure}

\begin{figure}[ht]
	\centering
	\includegraphics[scale=0.7]{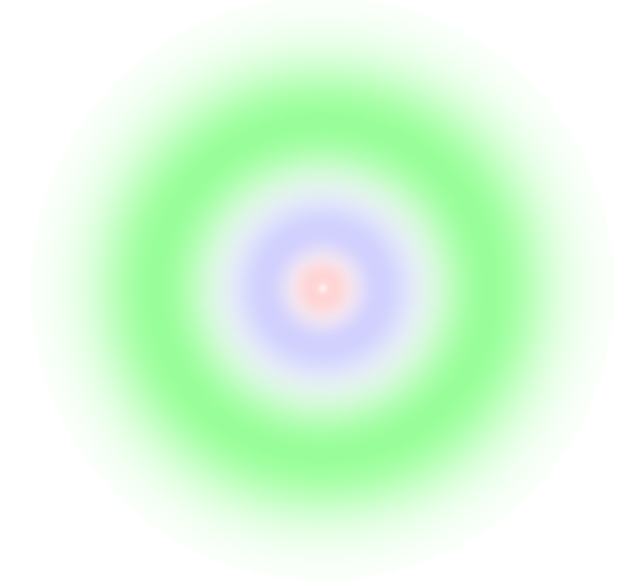}
	\caption{Planar section, passing through the center, of the energy density, in which the contributions $\rho_1$ (red), $\rho_2$ (blue) and $\rho_3$ (green) calculated in subsection~\ref{second} are shown, giving rise to a multimagnetic structure. We depict the case $\alpha=1$, $\beta=10$, $\gamma=30$.}
	\label{threeB}
	%\end{center}
\end{figure}

As our second example, we abandon the standard core of the previous one, and take $P^{(1)}=|\phi^{(1)}|^{-\alpha}$, implying a self-interaction, as well as $P^{(2)}=|\phi^{(1)}|^{-\beta}$ and $P^{(3)}=|\phi^{(2)}|^{-\gamma}$. This means that we are now using $\alpha=\alpha^1_1$, $\beta=\alpha^2_1$, and $\gamma=\alpha^3_2$, with all other parameters being zero. This combination of parameters results in a triple shell structure distinct of the previous one, now with a hollow center and almost empty shells in between the red and blue and the blue and green portions of the multimagnetic structure. In this example, the system of first order equations takes the form
\begin{subequations}\label{3d}
\begin{align}
&H^{(1)'}=\left(\frac{1}{H^{(1)}}\right)^{\alpha}\frac{1-{K^{(1)}}^2}{r^{2}},\\
&K^{(1)'}=-H^{(1)}K^{(1)}\left(H^{(1)}\right)^{\alpha},
\end{align}
\end{subequations}
and
\begin{subequations}\label{3e}
\begin{align}
&H^{(2)'}=\left(\frac{1}{H^{(1)}}\right)^{\beta}\frac{1-{K^{(2)}}^2}{r^{2}},\\
&K^{(2)'}=-H^{(2)}K^{(2)}\left(H^{(1)}\right)^{\beta},
\end{align}
\end{subequations}
and
\begin{subequations}\label{3f}
\begin{align}\label{BPS6}
&H^{(3)'}=\left(\frac{1}{H^{(2)}}\right)^{\gamma}\frac{1-{K^{(3)}}^2}{r^{2}},\\
&K^{(3)'}=-H^{(3)}K^{(3)}\left(H^{(2)}\right)^{\gamma}.
\end{align}
\end{subequations}
Closed form solutions have not been discovered for any of the functions in this example, but we may solve these equations numerically. Once again, we find that the $H^{(n)}$ and $K^{(n)}$ present profiles that are qualitatively similar to those of the standard monopole solution. These solutions, as well as the individual contributions to the energy density, are represented in Figs.~\ref{profile4},~\ref{profile5},~\ref{profile6}, and \ref{threeB}, for $\alpha=3$, $\beta=10$ and $\gamma=30$, to produce the desired result, well distinct from the previous case; compare Figs. \ref{three} and \ref{threeB}. The total energy is again equal to $12\pi$.

\subsection{Third Model}\label{ex3}

We may also consider another possibility, which can be achieved by taking $P^{(2)}$ and  $P^{(3)}$ as functions of $|\phi^{(1)}|$ alone. We proceed in a way similar to the previous two cases, but now we consider $P^{(1)}=1$, $P^{(2)}=|\phi^{(1)}|^{-\alpha}$ and $P^{(3)}=|\phi^{(1)}|^{-\beta}$. This leads to equations~\eqref{BPS1} and~\eqref{3b}, coupled to the additional pair of equations
\bes\label{3g}
\begin{align}
&H^{(3)'}=\left(\frac{1}{H^{(1)}}\right)^{\beta}\frac{1-{K^{(3)}}^2}{r^{2}},\\
&K^{(3)'}=-H^{(3)}K^{(3)}\left(H^{(1)}\right)^{\beta}.
\end{align}
\ees 

\begin{figure}[ht]
	\centering
	\includegraphics[scale=0.21]{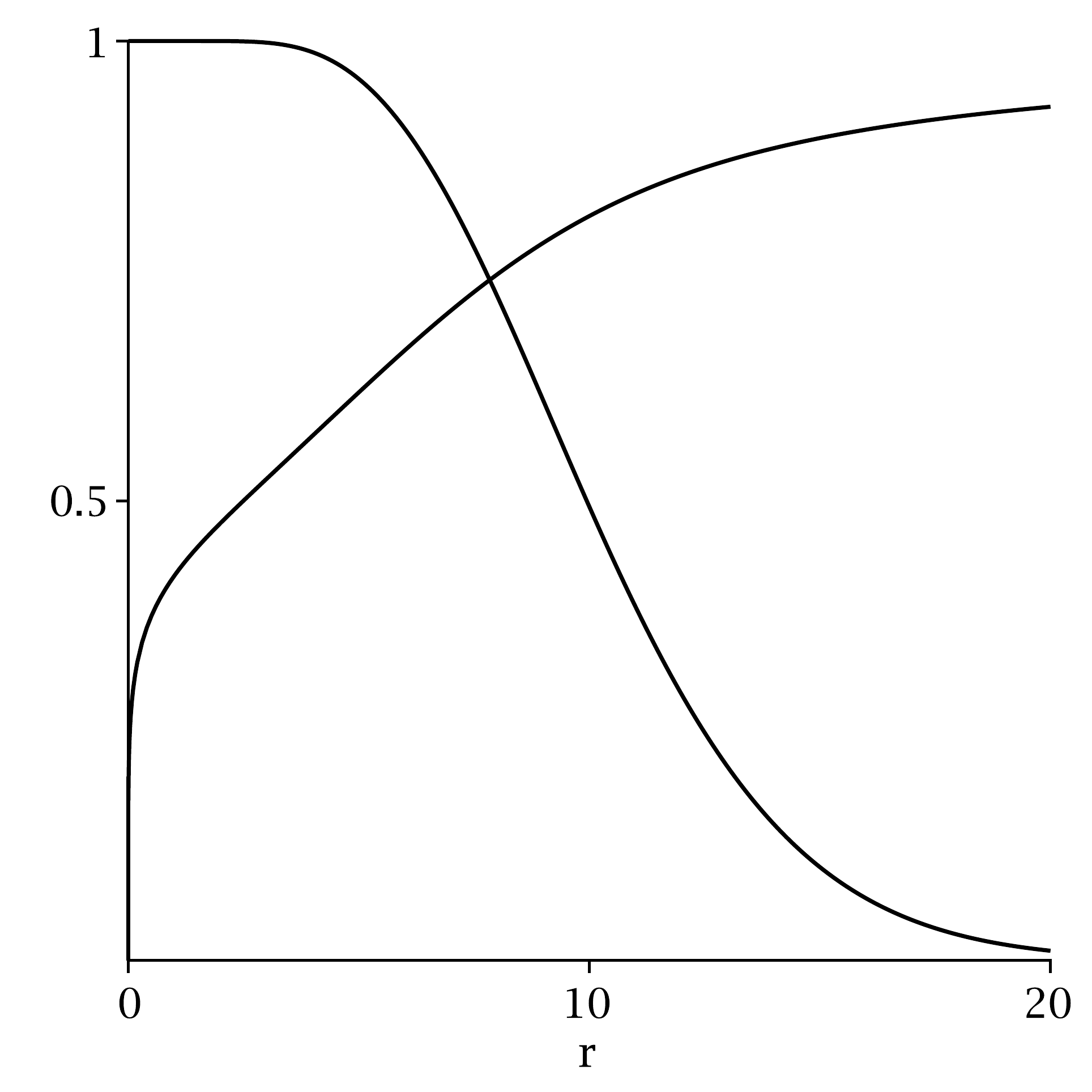}
	\includegraphics[scale=0.21]{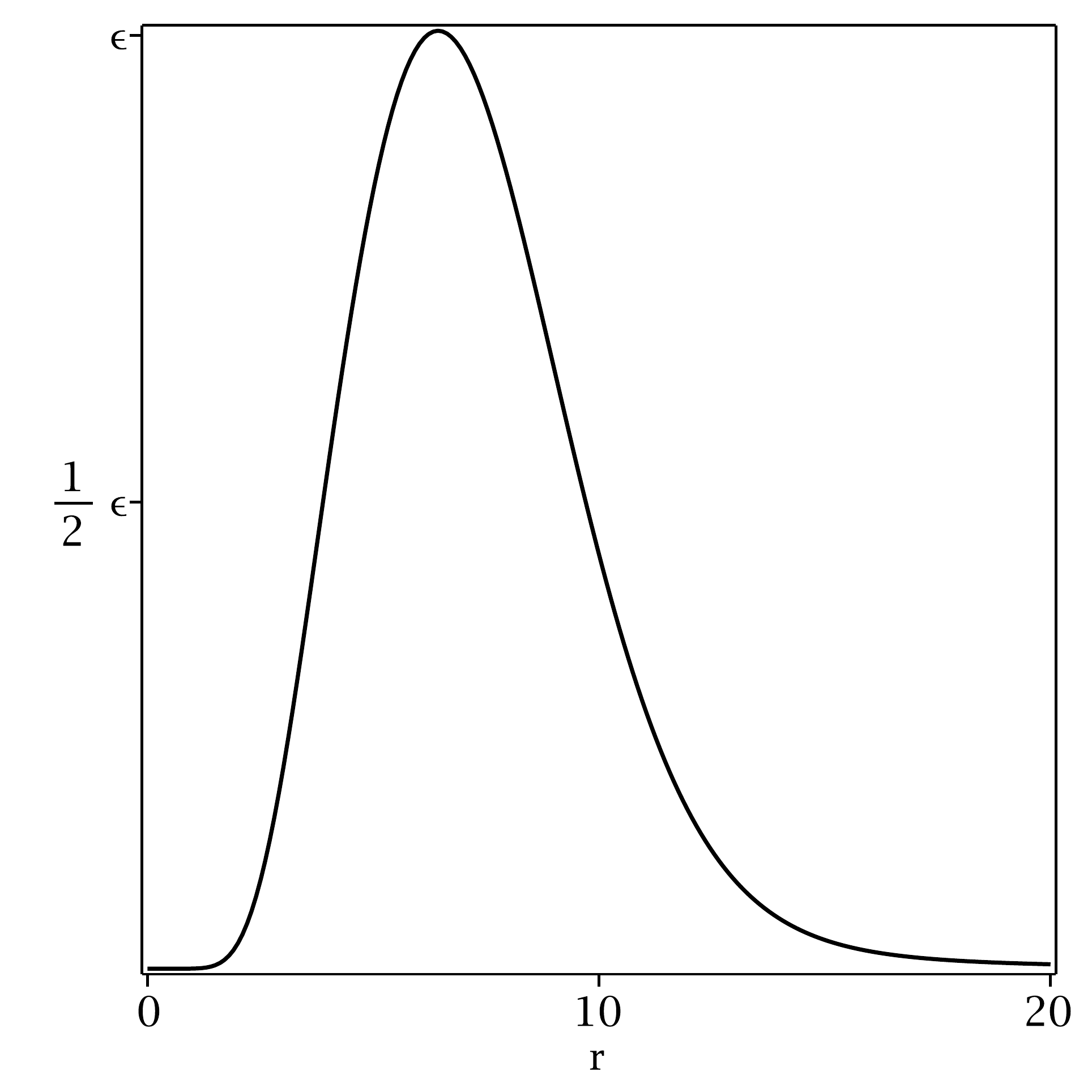}
	\caption{In the left panel, we show the solutions $H^{(3)}(r)$ (ascending line) and $K^{(3)}(r)$ (descending line) of~\eqref{3g} for $\beta=12$. In the right panel, we show the contribution $\rho_3$, calculated from~\eqref{rhoa}, relative to this sector. Here, $\epsilon=0.0025$.}
	\label{ss}
	%\end{center}
\end{figure}

\begin{figure}[ht]
	\centering
	\includegraphics[scale=0.7]{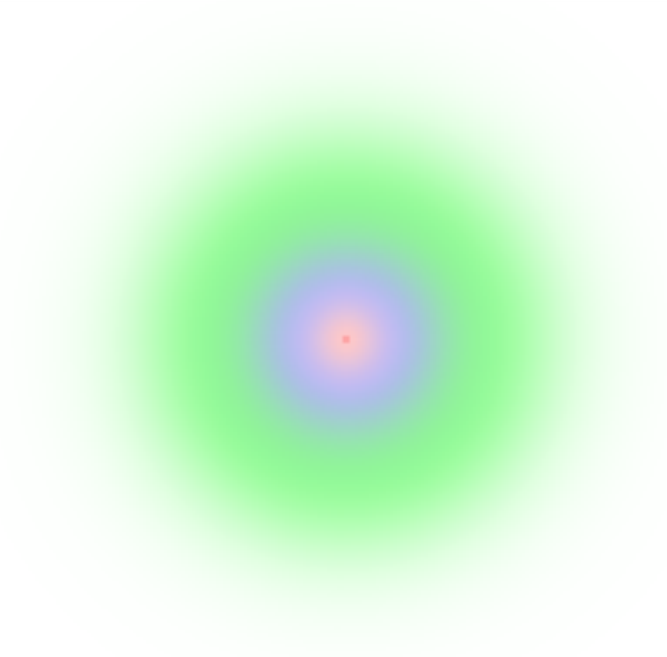}
	\caption{Planar section, passing through the center, of the energy density, in which the contributions $\rho_1$ (red), $\rho_2$ (blue) and $\rho_3$ (green) calculated in subsection~\ref{ex3} are shown, giving rise to a multimagnetic structure. We depict the case $\alpha=3$ and $\beta=12$.}
	\label{singlesource}
	%\end{center}
\end{figure}

This modification does not change the four other equations, the functions $H^{(1)}, K^{(1)}, H^{(2)}$ and $K^{(2)}$ are the same discussed in subsection~\ref{ex1}. The substitution $H^{(1)}(r)=\coth(r) - 1/r$ in~\eqref{3g} gives rise to a system of ODEs that can be solved for $H^{(3)}$ and $K^{(3)}$. Their features are shown in Fig.~\ref{ss}, for $\beta=12$. The multimagnetic structure that arises from the solution of the equations of motion is depicted in Fig.~\ref{singlesource}, showing another way to localize the red, blue and green shells of the triple configuration; compare the three Figs. \ref{three}, \ref{threeB} and \ref{singlesource}.

%%%%%%%%%%%%%%%%%%%%%
\subsection{Fourth Model}\label{ex4}

As a final example of the $N=3$ class of models, let us consider another example, in which $P^{(1)}=1$, $P^{(2)}=|\phi^{(1)}|^{-\alpha}$ and $P^{(3)}=(|\phi^{(1)}||\phi^{(2)}|)^{-\beta}$. This model is different from the previous examples in that both the $\phi^{(1)}$ and $\phi^{(2)}$ fields appear explicitly in the third sector, with first order equations that are given by 
\bes\label{3h}
\begin{align}
&H^{(3)'}=\left(\frac{1}{H^{(1)}H^{(2)}}\right)^{\beta}\frac{1-{K^{(3)}}^2}{r^{2}},\\
&K^{(3)'}=-H^{(3)}K^{(3)}\left(H^{(1)}H^{(2)}\right)^{\beta}.
\end{align}
\ees 

\begin{figure}[ht]
	\centering
	\includegraphics[scale=0.21]{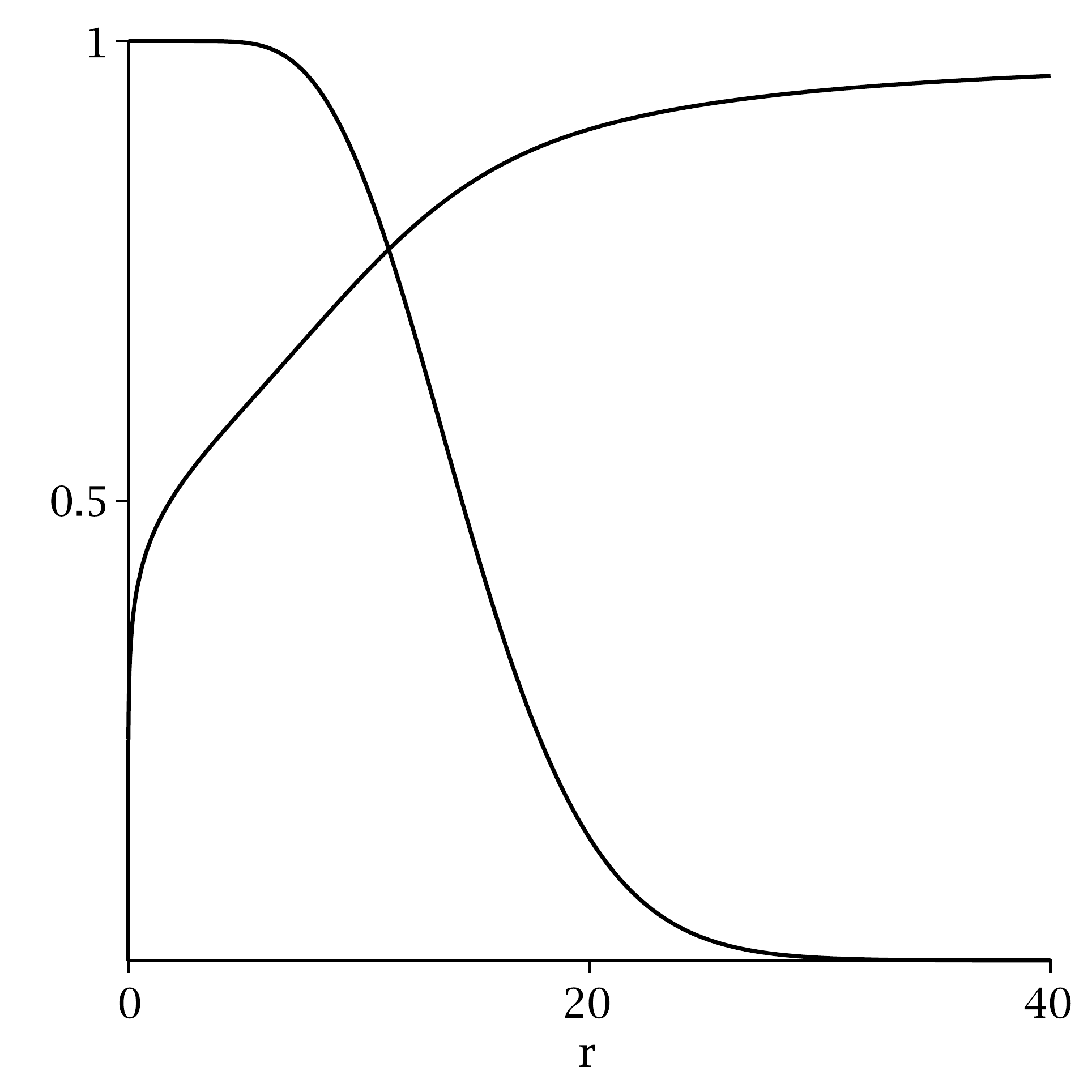}
	\includegraphics[scale=0.21]{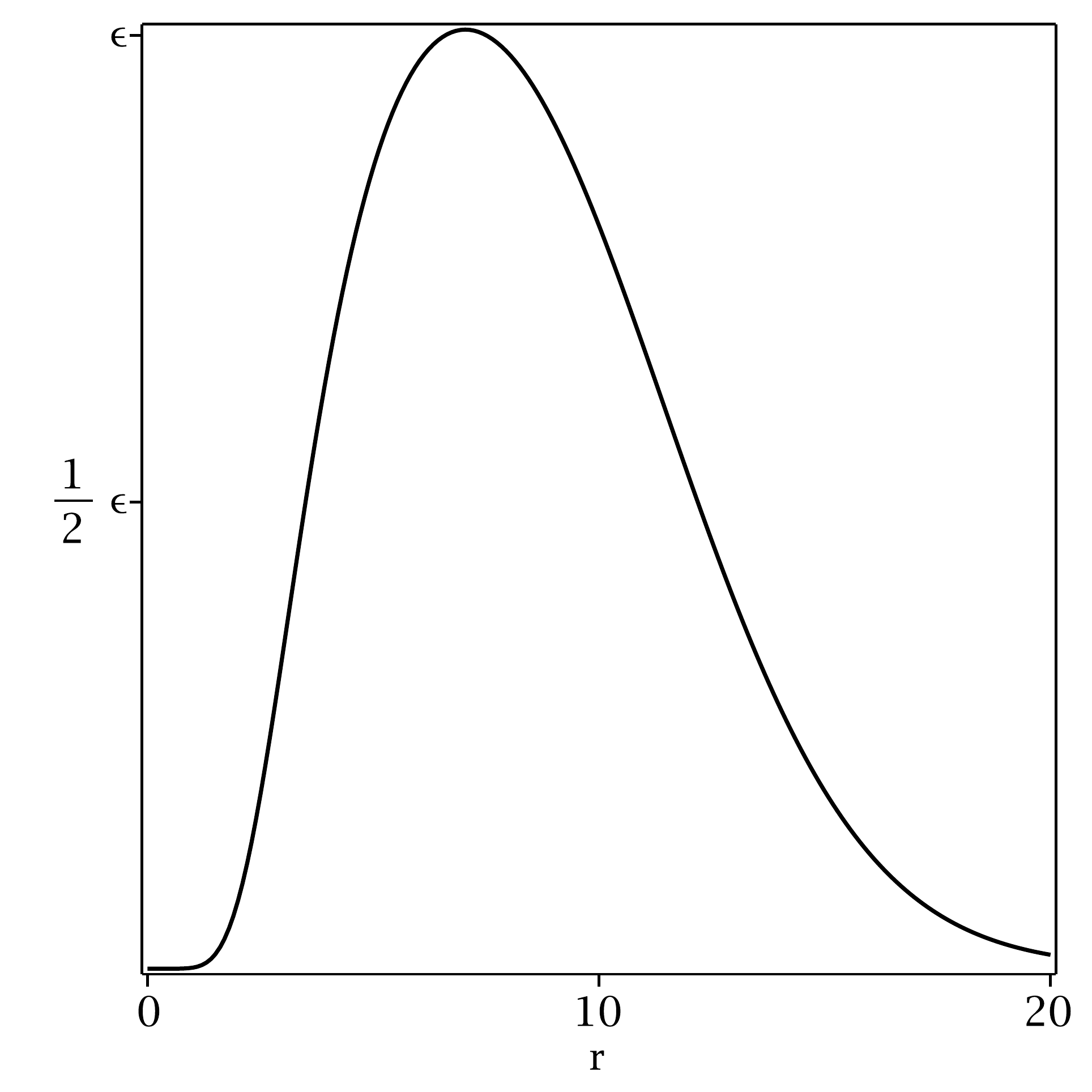}
	\caption{In the left panel, we show the solutions $H^{(3)}(r)$ (ascending line) and $K^{(3)}(r)$ (descending line) of~\eqref{3h} for $\alpha=3$ and $\beta=10$. In the right panel, we show the contribution $\rho_3$, calculated from~\eqref{rhoa}, relative to this sector. Here, $\epsilon=0.0036$.}
	\label{3N}
	%\end{center}
\end{figure}

\begin{figure}[ht]
	\centering
	\includegraphics[scale=0.6]{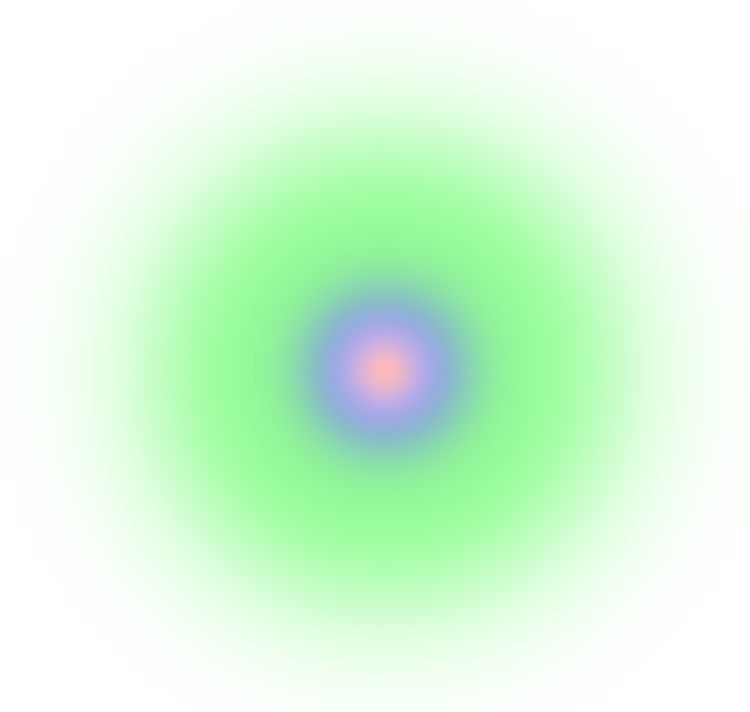}
	\caption{Planar section, passing through the center, of the energy density, in which the contributions $\rho_1$ (red), $\rho_2$ (blue) and $\rho_3$ (green) calculated in subsection~\ref{ex4} are shown, giving rise to a multimagnetic structure. We depict the case $\alpha=3$ and $\beta=10$.}
	\label{4N}
	%\end{center}
\end{figure}

These equations are coupled to~\eqref{BPS1} and~\eqref{3b}, and we can plug in the results for $H^{(1)}$ and $H^{(2)}$ to solve~\eqref{3h}. This has been done for the case $\alpha=3$ and $\beta=10$. The solutions of~\eqref{3h} are displayed in Fig.~\ref{3N} and the multimagnetic structure associated to this case is shown in Fig.~\ref{4N}.

%%%%%%%%%%%%%%%%%%%%%%%%%%%%%%% N=4 %%%%%%%%%%%%%%%%%%%%%%%%%%%%%%%%%%%%%%%%%%%

\section{MODELS WITH N=4}\label{N4}

Let us now investigate another system, resulting in the group $\rm{SU(2)}\times\rm{SU(2)}\times \rm{SU(2)}\times \rm{SU(2)}$. This requires that we have four pairs of first order differential equations. The simplest way to achieve this is by adding a fourth pair of fields to the systems described in the previous section, and we investigate some distinct possibilities below. 

\subsection{First Model}\label{third}
 As a first example, we take $P^{(1)}$, $P^{(2)}$ and $P^{(3)}$ as in our very first model of the previous Section and, additionally, choose $P^{(4)}=|\phi^{(3)}|^{-\gamma}$. We are then led to equations~\eqref{BPS1},~\eqref{3b} and~\eqref{3c}, with the additional pair of first order equations
\bes\label{4a}
\begin{align}
&H^{(4)'}=\left(\frac{1}{H^{(3)}}\right)^{\gamma}\frac{1-{K^{(4)}}^2}{r^{2}},\\
&K^{(4)'}=-H^{(4)}K^{(4)}\left(H^{(3)}\right)^{\gamma}.
\end{align}
\ees

\begin{figure}[ht]
	\centering
	\includegraphics[scale=0.21]{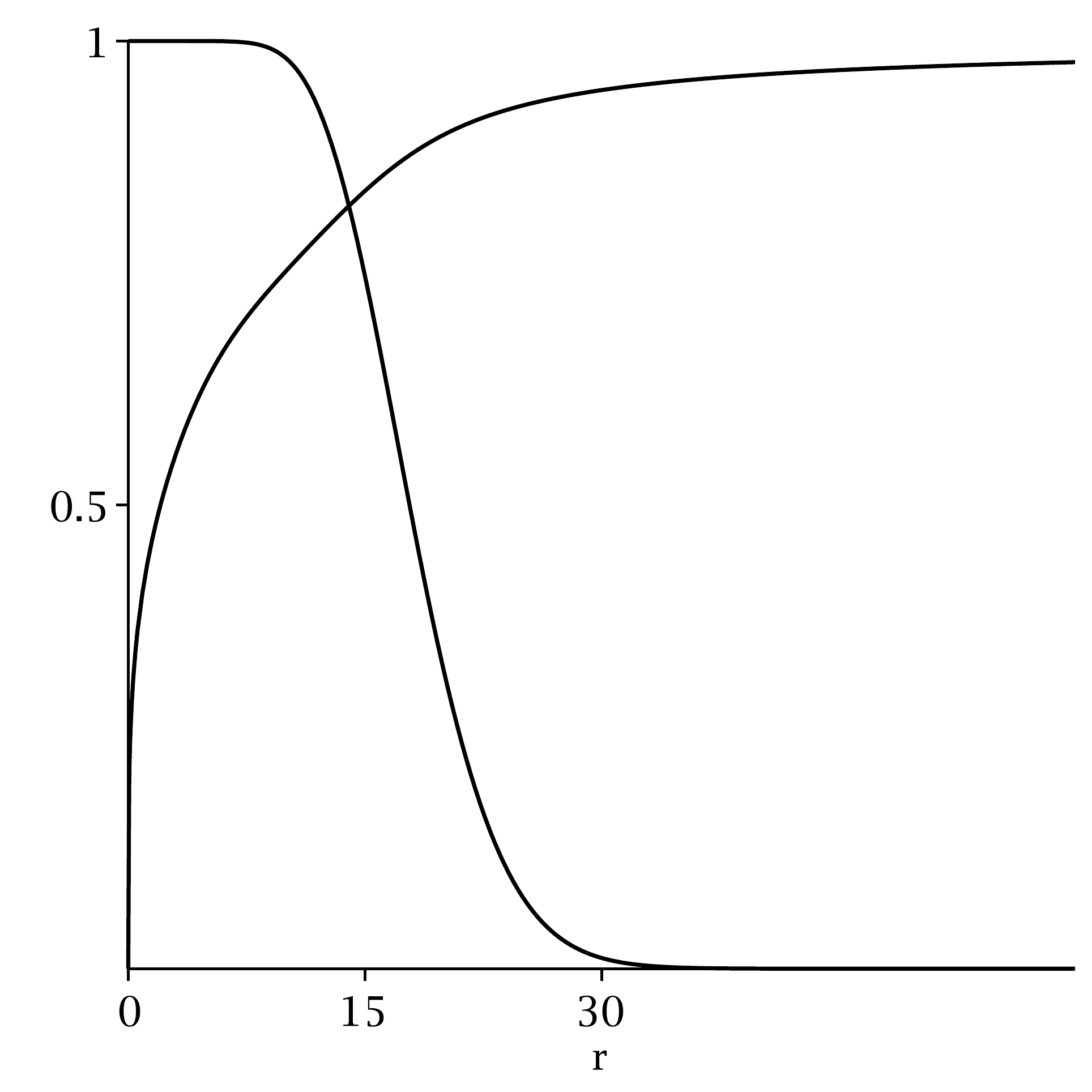}
	\includegraphics[scale=0.21]{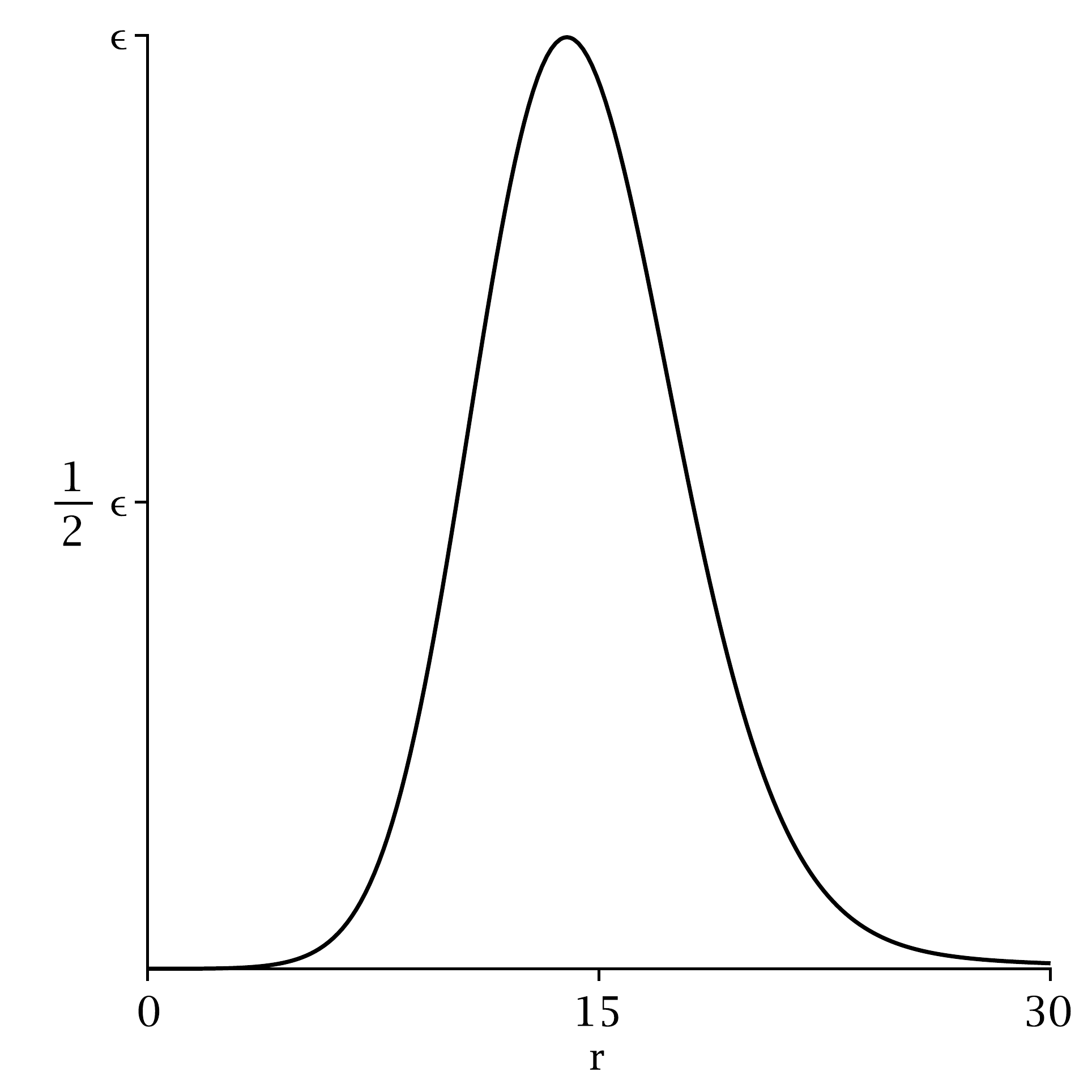}
	\caption{In the left panel, we show the graph of the functions $H^{(4)}$ (ascending line) and $K^{(4)}$ (descending line) calculated from~\eqref{4a} which, together with the functions obtained in subsection~\ref{ex1}, form a solution of the equations of motion. In the  right panel, we show the contribution $\rho_4$, calculated from~\eqref{rhoa}, relative to this sector. Both plots correspond to the choices $\alpha=3$, $\beta=10$, $\gamma=20$ of the model's parameters. Here, $\epsilon=0.0005$.}
	\label{densc}
\end{figure}

Since a solution for $H^{(3)}$ has already been found numerically in subsection~\ref{ex1}, we may plug this result in the equations above, and then solve them numerically. The result for the case $\alpha=3$, $\beta=10$, $\gamma=20$, along with the contribution $\rho_4$ to the energy density, is shown in Fig.~\ref{densc}. We see that $\rho_4$ attains its maximum at $r\simeq 13.9$, and contributes to the energy with another $4\pi$ factor, raising its value to $16\pi$.

\begin{figure}[ht]
	\centering
	\includegraphics[scale=0.7, trim={0cm 0cm 0cm 0cm}, clip]{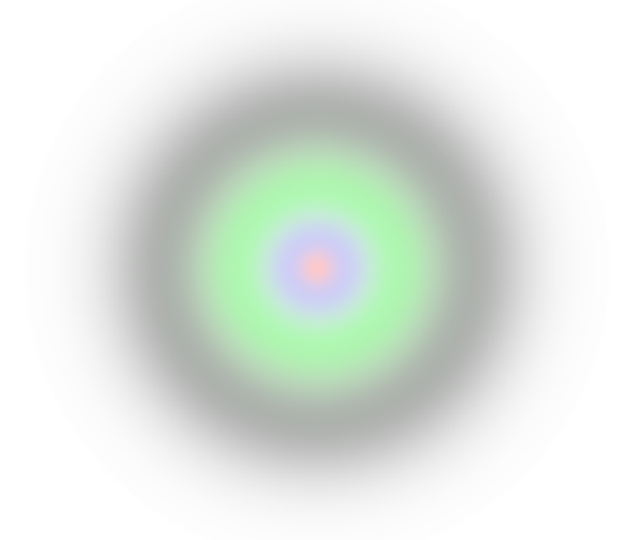}
	\caption{Planar section, passing through the center, of the four contributions $\rho_1$ (red), $\rho_2$ (blue), $\rho_3$ (green) and  $\rho_4$ (black) relative to the model presented in subsection~\ref{third}.}
	\label{fourA}
\end{figure}

Moreover, we depict in Fig.~\ref{fourA} the multimagnetic structure which results from the superposition of the four subsystems considered, with the black color used to describe the energy density $\rho_4$ of the new ${\rm SU(2)}$ symmetry added to the system; compare with Fig. \ref{three}.
\vspace{40pt}

\subsection{Second Model}\label{fourth}
\begin{figure}[ht]
	\centering
	\includegraphics[scale=0.7, trim={0cm 0cm 0cm 0cm}, clip]{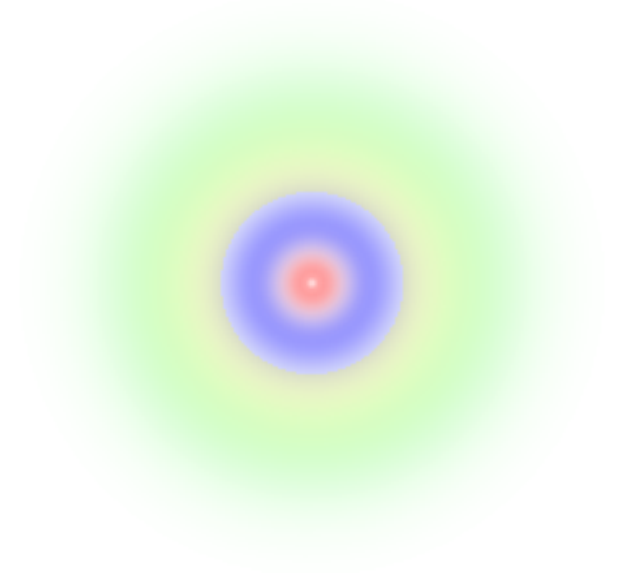}
	\caption{Planar section, passing through the center, of the four contributions $\rho_1$ (red), $\rho_2$ (blue), $\rho_3$ (yellow) and  $\rho_4$ (green) to the energy density relative to the model presented in subsection~\ref{fourth}.}
	\label{fourb}
\end{figure}

As another example, we present a solution with four shells. To achieve this, we take $P^{1}$, $P^{2}$ and $P^{3}$ as in subsection~\ref{third}, as well as $P^{(4)}=|\phi^{(3)}|^{-\zeta}$. This case results in a system of equations comprised by~\eqref{3d}, \eqref{3e} and~\eqref{3f}, which are now coupled with the additional pair of equations
\bes \label{4b}
\begin{align}
&H^{(4)'}=\left(\frac{1}{H^{(3)}}\right)^{\zeta}\frac{1-{K^{(4)}}^2}{r^{2}},\label{d1}\\
&K^{(4)'}=-H^{(4)}K^{(3)}\left(H^{(3)}\right)^{\zeta}\label{d2}.
\end{align}
\ees
The equations have again been solved through numerical procedures, and the results can be seen in Fig.~\ref{densd} below. The energy density $\rho_4(r)$ has a peak at $r\approx 8.2$. We then present in Fig.~\ref{fourb} the multimagnetic structure that arises from the superposition of the four subsystems considered.

\begin{figure}[ht]
	\centering
\includegraphics[scale=0.21]{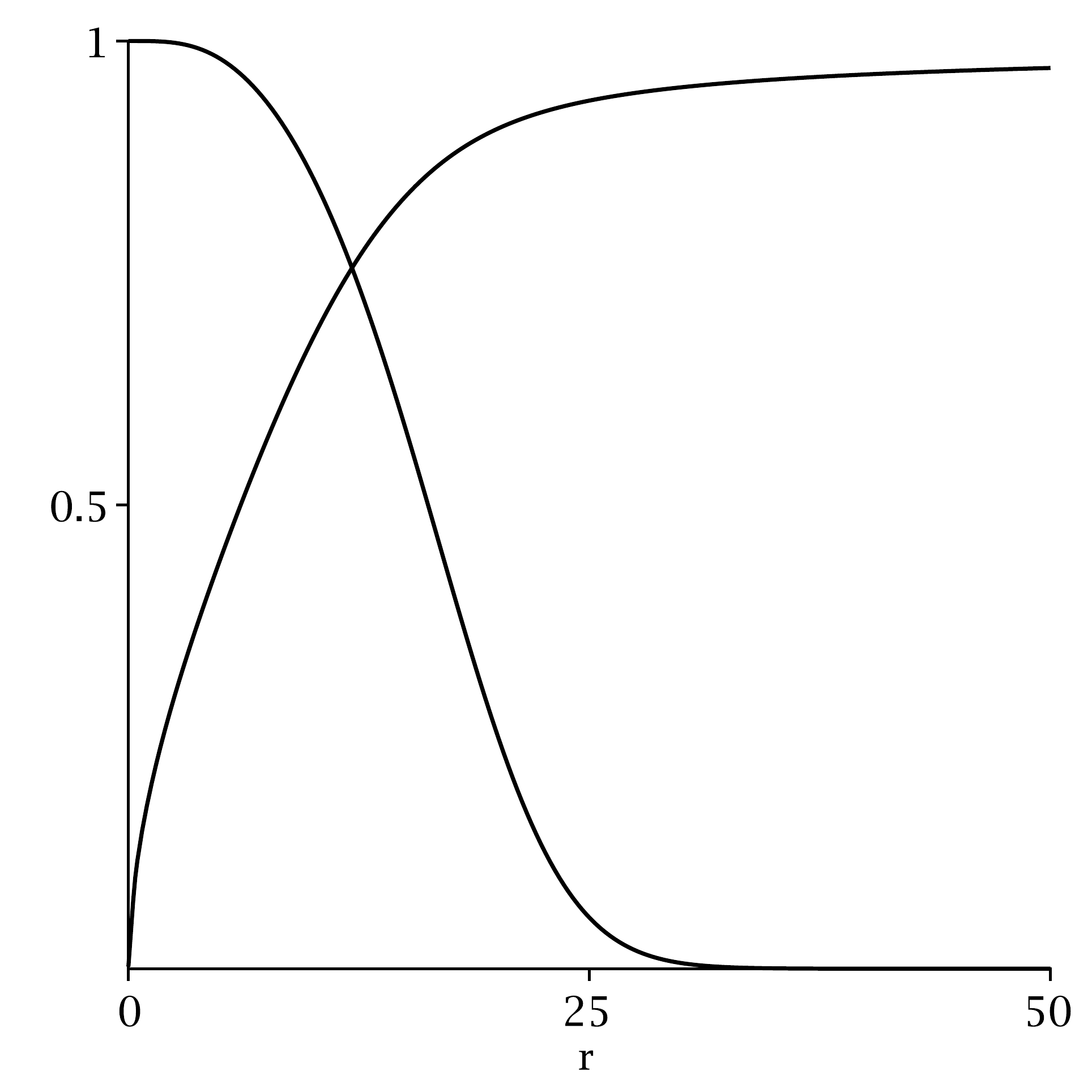}
\includegraphics[scale=0.21]{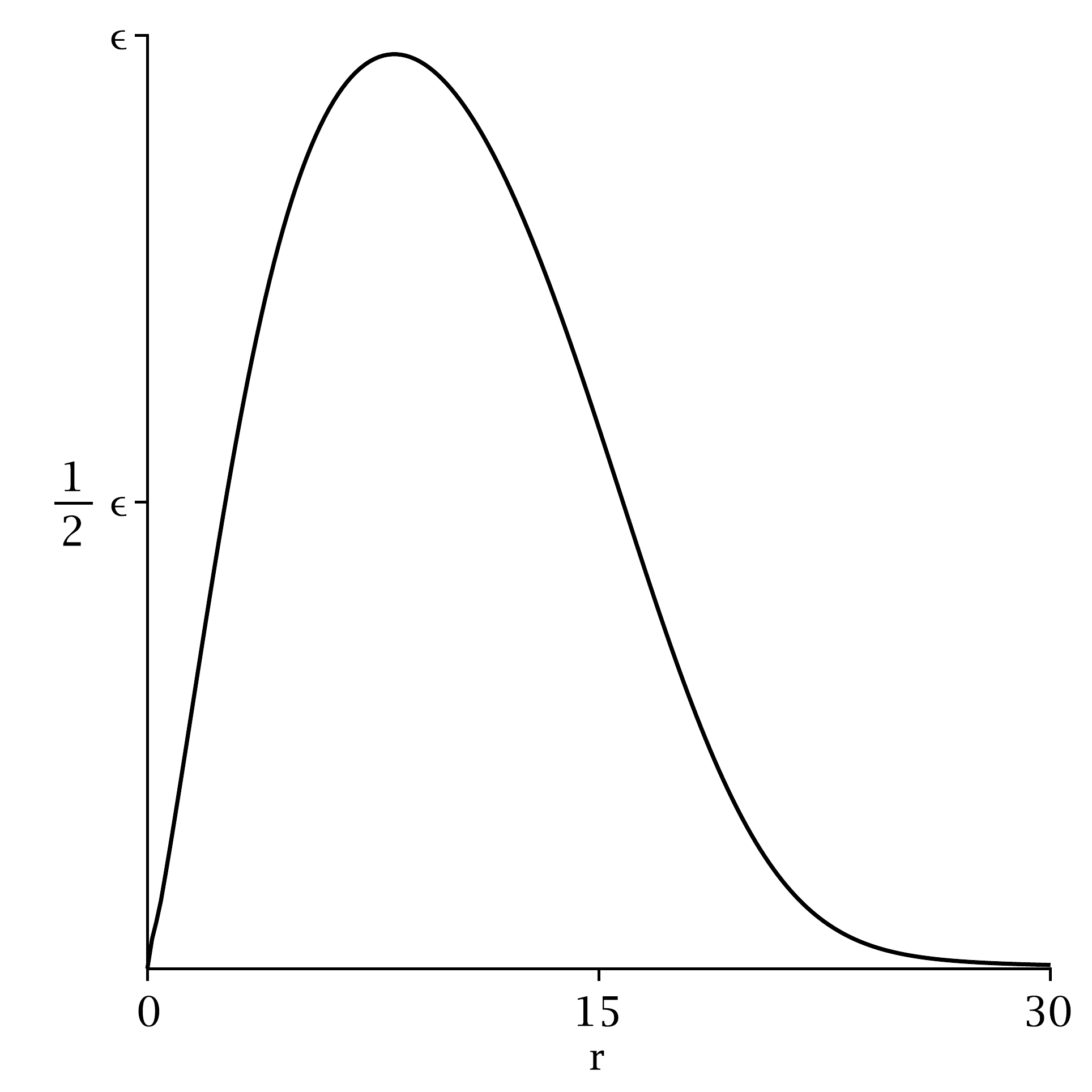}
	\caption{In the left panel, we show the graph of the functions $H^{(4)}$ and $K^{(4)}$ which solve~\eqref{4b} and, together with the functions calculated in subsection~\ref{second}, form a solution of the equations of motion. In the  right panel, we show the contribution $\rho_4$ to the energy density. Both plots correspond to the choices $\alpha=1$, $\beta=10$, $\gamma=30$, $\zeta=8$ of the model's parameters. Here, $\epsilon=0.0006$.}
	\label{densd}
\end{figure}

In the case with $N=4$ there are several other possibilities, similar to the case of $N=3$ which we have investigated above.

\section{DISCUSSION}\label{disc}

In this work we studied the presence of magnetic monopoles in a relativistic model with a modified Lagrangian that is invariant under $N$ copies of the $\rm{SU(2)}$ symmetry, that is, $\rm{SU_{(1)}(2)}\times\rm{SU_{(2)}(2)}\times\cdots \times \rm{SU_{(N)}(2)}$. We developed a general investigation leading to the presence of first order differential equations that solve the equations of motion and ensure the presence of minimum energy configurations. We then discussed some examples with $\rm N=3$ and $\rm N=4$, with shells on top of a standard monopole core and in other cases, giving rise to novel onionlike structures. We believe that these multimagnetic structures may find applications in several areas of nonlinear science. In particular, the solutions generalize the bimagnetic structures considered before in \cite{bimag}, and they may find applications in several different scenarios in condensed matter \cite{Estrader,EstraderII,PRL20,nano}. 

The distinct multimagnetic structures that we have described in the present work may respond differently under the action of an external magnetic field, for instance, so they may be identified via a magnetic probe and, in this sense, be of application in several distinct scenarios of current interest in nonlinear physics in general. Instead of the replication of the $\rm{SU(2)}$ symmetry, we can also chose other models with gauge groups of the form $G_1\times...\times G_N$, where the $G_k$ are themselves non Abelian gauge groups. One particularly interesting example is found for the gauge group $\rm{SO(10)}$. This particular grand unification theory has been studied for quite some time~\cite{SO(10), SO(10)II} and is considered a relatively simple and promising unification model~\cite{GUT, SO10}. This gauge group may undertake symmetry breaking to $\rm{SU(3)}\times\rm{SU(2)}\times\rm{SU(2)}\times\rm{U(1)}$ as an intermediary step before reduction to the gauge group of the standard model~\cite{gonzalo}. Thus, we could take, for example, an $\rm{SU(3)}$ core with two shells provided by the symmetry breaking from the $\rm{SU(2)}\times\rm{SU(2)}$ subgroup. Although mathematically more challenging, this case is, in principle, a straightforward extension of the examples provided here. In the same way, we could also consider the SU(3) and SU(5) gauge groups, since they also support monopoles \cite{PRD,PRL} and may be used with distinct motivations. Another line of current interest concerns the study of the magnetic monopoles found in the present work interacting with gravity, to see how they change the standard scenario of black holes in magnetic monopoles \cite{Wei}. As we can see, the underlying principle leading to the solutions studied in this work may lead to a wide range of multimagnetic monopoles, differing by the gauge group and by the many possible choices for the functions that modify the standard model. We hope that the above results will foster new investigation on magnetic monopoles.

The main results of the present work motivate us to think of diminishing the spatial dimension, investigating, for instance, vortices in planar systems. In this case, the basic $\rm U(1)$ symmetry can be enhanced to become $\rm U_1(1)\times U_2(1)\times\cdots \times U_N(1)$, bringing into action a system with $\rm N$ order parameters. This study can be developed in direct connection with the works \cite{multilayered,prl2,prl22,prl3}, and is presently under consideration. It is also of interest to the case of multi-component Gross-Pitaevskii equations to describe solutions in the case of spin-1 and spin-2 Bose-Einstein condensates; see, e.g., Ref. \cite{PRep} and references therein. We can also go further down and consider lineal systems, and study kinks in one spatial dimension. Here we can follow the lines of \cite{multikink}, where we studied kinks in a model with $\rm Z_2\times Z_2$ symmetry. Extension to the case of the $\rm Z_2\times Z_2\times\cdots\times Z_2$ symmetry can follow the present investigation, and is also under consideration. We hope to report on some of these issues in the near future.

Another line of investigation is related to the recent study in Ref. \cite{ahep}, where the authors found magnetic monopole solution in a multivector boson theory in which a replication of the basic SU(2) symmetry is also considered. The construction of the magnetic monopole described in \cite{ahep} is based on the dimensional deconstruction mechanism \cite{DD} and the Higgsless theory \cite{HL}, and this is different from the procedure which we described in the present work, where we focus on stable minimum energy configurations with multimagnetic properties. We would also like to add that the multimagnetic structures which appeared in this work may be constructed under the guidance of the recent advances on the study of multishelled hollow micro-nanostructures with triple or more shells; see, e.g., Refs. \cite{AAA,BBB} and references therein. In these investigations, the main motivation included potential applications mainly in the fields of energy conversion and storage, sensors, photocatalysis, and drug delivery, but the above results may foster investigations concerning the presence of shells with distinct magnetic properties. 
  
\acknowledgements{The work is supported by the Brazilian agencies Coordena\c{c}\~ao de Aperfei\c{c}oamento de Pessoal de N\'ivel Superior (CAPES), grant No 88887.485504/2020-00 (MAL), Conselho Nacional de Desenvolvimento Cient\'ifico e Tecnol\'ogico (CNPq), grants No. 303469/2019-6 (DB) and No. 404913/2018-0 (DB), and by Paraiba State Research Foundation (FAPESQ-PB) grant No. 0015/2019.}
%%%%%%%%%%%%%%%%%%%%%%%%%%%%%%%%%

\end{document}